\documentclass[a4paper,11pt]{article}

\usepackage[utf8]{inputenc}
\usepackage{jheppub} 

\usepackage{slashed}
\usepackage{epsfig}
\usepackage{url}
\usepackage{dcolumn}
\usepackage{bm}
\usepackage[usenames ,dvipsnames]{xcolor}
\usepackage{slashed}
\usepackage{graphicx,color}
\usepackage{url}
\usepackage{multirow,array}
\usepackage{float}
\usepackage{fancyhdr}
\usepackage{indentfirst}
\usepackage{cancel}
\usepackage{tabularx}
\usepackage{simplewick}
\usepackage[normalem]{ulem}
\newcommand{\st}{\scriptsize}

\title{\boldmath Gravitational Waves from Global Cosmic Strings and Cosmic Archaeology}

\author[a]{Chia-Feng Chang and}
\author[a]{Yanou Cui}

\affiliation[a]{Department of Physics and Astronomy, University of California, Riverside, CA 92521, USA}

\emailAdd{chiafeng.chang@email.ucr.edu}
\emailAdd{yanou.cui@ucr.edu}

\abstract{
Global cosmic strings are predicted in many motivated extensions to the Standard Model of particle physics, with close connections to axion dark matter physics. Recent studies suggest that, although subdominant relative to Goldstone emission, gravitational wave (GW) signals from global strings can be detectable with current and planned GW detectors such as LIGO, LISA, DECIGO/BBO, ET/CE and AEDGE/AION, as well as pulsar timing arrays such as PPTA, NANOGrav and SKA. This work is an extensive, updated study on GWs from a global cosmic string network, taking into account of the most recent developments related to the subject. The main analysis is based on the analytical Velocity-dependent One-Scale (VOS) model calibrated with recent simulation results, which provides a generic protocol for such calculations with details given. We also demonstrate how the GW signal can be influenced with variations to the baseline model: this includes considering the uncertainties of model parameters and the potential deviation from the conventional VOS model prediction (i.e.~the scaling behavior) as suggested by some of the recent simulation results. Furthermore, we investigated in detail the effect of a non-standard cosmology (e.g.~early matter domination or kination) or new particle species on the GW signals from global strings. We demonstrate that the frequency spectrum of GW background from global cosmic strings can be used to probe the cosmic history prior to the Big Bang nucleosynthesis (BBN) (i.e.~the primordial dark age) up to a temperature of $T\sim10^8$ GeV.}

\begin{document} 
\maketitle
\flushbottom

\section{Introduction}
\label{sec:intro}
The detection of gravitational waves (GW) by the LIGO/Virgo collaboration \cite{Evans:2016mbw,Abbott:2016,Abbott:2017mem} opens up a new observational window into the cosmos, and offers unprecedented opportunities to probe fundamental physics beyond the Standard Model (SM). The presence of a cosmologically generated stochastic GW background (SGWB) is highly motivated and has been actively searched for/studied by the LIGO and LISA collaborations \cite{TheLIGOScientific:2016wyq,LIGOScientific:2019vic,Audley:2017drz,Bartolo:2016ami}. Although still being investigated, the intriguing stochastic signal recently reported by the NANOGrav collaboration \cite{Arzoumanian:2018saf,Arzoumanian:2020vkk,Arzoumanian:2021teu} has been shown to be possibly explained by a SGWB of cosmic origin \cite{Ellis:2020ena,Blasi:2020mfx,DeLuca:2020agl,Buchmuller:2020lbh,Vagnozzi:2020gtf,Kohri:2020qqd,Ramberg:2020oct,Blanco-Pillado:2021ygr}. Furthermore, the detection of a cosmogenic SGWB can potentially address many long-standing questions in particle physics and cosmology (e.g. \cite{Grojean:2006bp,Schwaller:2015tja,Cui:2017,Cui:2018,Caldwell:2018giq,Chang:2019mza,Gouttenoire:2019rtn,Cui:2019kkd,Buchmuller:2019gfy,Dror:2019syi,Dunsky:2019upk,Blasi:2020wpy,Machado:2019xuc}), and allows us to probe very early stages of the Universe.

Among the known cosmological sources of SGWB (see review \cite{Caprini:2018mtu}), cosmic strings stand out as one that can yield strong signals over a wide frequency range due to continuous emission throughout a long period of time. Cosmic strings are one-dimensional, topologically stable objects that are generically predicted by many theoretical extensions of the Standard Model of particle physics, e.g., field theories with a spontaneously broken $U(1)$ symmetry (gauge or global) \cite{Kibble:1976sj,Nielsen:1973cs,Vachaspati:1984dz,Vilenkin:2000jqa,King:2020hyd,Huang:2020bbe,Huang:2020mso}, and the fundamental and/or composite strings in superstring theory \cite{Copeland:2003bj,Dvali:2003zj,Polchinski:2004ia,Jackson:2004zg,Tye:2005fn}. After formation, the strings quickly evolve towards a scaling regime where the string network consists of a few Hubble-length long strings per horizon volume, along with more copious loops formed by long string intersections. The loops then oscillate and radiate energy in the form of GWs and/or other particles until they decay away. 
Most literature on GW signatures from cosmic strings have been focused on those sourced by local strings or superstrings which typically can be described by Nambu-Goto (NG) action. 
In contrast, a global string network as a potential source of GWs has been largely ignored since by naive estimate GW radiation would be overwhelmed by Goldstone emission which occurs with a much larger rate. Very recently, inspired by its intimate connections to axion dark matter physics, significant progress has been made in simulating global topological defects and on the GW signals originated from it \cite{Gorghetto:2018myk,Buschmann:2019icd,Gorghetto:2020qws,Figueroa:2020lvo,Gorghetto:2021fsn}. With a semi-analytical approach based on the Velocity-dependent One-Scale (VOS) model, our earlier work \cite{Chang:2019mza} demonstrated that the GW signal from global strings, albeit notably smaller than that from its NG string counterpart, can be within reach of future GW experiments such as LISA \cite{Audley:2017drz,Bartolo:2016ami}, AEDGE \cite{Bertoldi:2019tck}, DECIGO and BBO \cite{Yagi:2011wg}. Such a positive prospect of detection has been confirmed by simulation-based work \cite{Gorghetto:2021fsn,Figueroa:2020lvo}, although details differ which will be addressed in this work. 

The frequency spectrum of the SGWB from a cosmic string network can also serve as a powerful tool to probe the very early cosmic history that is not accessible by existing means. The $\Lambda$CDM cosmology was established based on precise measurements of electromagnetic radiation over different frequency ranges with a variety of experiments. A simple extrapolation of $\Lambda$CDM cosmology back in time suggests that the Universe is radiation dominated from the recent matter-radiation equality all the way back to the end of inflation. This paradigm is supported by observing cosmic microwave background (CMB), the relic photons that started free traveling when the radiation temperature was about 0.3 eV. The success of BBN theory in predicting primordial abundances of light elements also provides evidence for a radiation dominated era up to $T\sim 5$ MeV. However, the hypothesis of radiation domination (RD) for epochs prior to BBN or at radiation temperature higher than $\sim 5$ MeV is yet to be experimentally tested. On the other hand, possibilities of non-standard pre-BBN cosmologies are well motivated by many grounds, such as dark matter \cite{Feng:2010gw,Lin:2019uvt}, axion physics \cite{DiLuzio:2020wdo,Marsh:2015xka}, baryogenesis \cite{Patel:2011th,Morrissey:2012db}, non-minimal inflation/reheating \cite{Ratra:1987rm,Linde:2007fr}, and string compactification \cite{Hindmarsh:1994re,Vachaspati:2015cma}. In particular, recently there has been an increased interest in the impact of non-standard cosmology on dark matter physics \cite{Erickcek:2017zqj,Redmond:2017tja, Erickcek:2020wzd}. The discovery of GWs leads to unprecedented opportunities to shed light on this mysterious pre-BBN primordial dark age \cite{Allen:1996vm,Boyle:2005se,Boyle:2007zx}. GWs are the only cosmic messengers that can travel freely throughout space-time since the Big Bang. They carry unique information about the earliest phases of the Universe’s evolution, beyond what can be assessed by observing EM radiations. Due to the continuous, potentially strong GW emissions from a string network throughout a long era of cosmic history, the SGWB frequency spectrum from cosmic strings is particularly appealing as a tool for looking back in time or \textit{cosmic archaeology} \cite{Cui:2017,Cui:2018, Caldwell:2018giq}. The application of this idea in the context of NG strings was recently proposed and studied in \cite{Cui:2017,Cui:2018}, based on a frequency-time (temperature) correspondence. Cosmic archaeology with global string induced GWs was only briefly discussed in \cite{Chang:2019mza}, which we will explore in great detail in this update.

In this work, we aim at an extensive study of SGWB signals originated from a global string network, and a comprehensive investigation into the potential new physics imprints in the pre-BBN Universe that can be detected with such a GW spectrum. Greater technical details are given, which may serve as a handy reference for future studies. Our primary approach is to use the analytic VOS model calibrated with simulation results (directly obtained for early times). Due to technical difficulties of simultaneously capturing physics at hierarchical scales, current simulations can only cover the evolution history of a global string network up to a few e-folds of Hubble expansion after the formation time. Thus, whether it is reliable to make a direct extrapolation of simulation results to late times (most relevant for observations today) requires further investigation. On the other hand, while VOS model for global strings are still being tested and needs to be calibrated with simulation data, the prediction for late times by the VOS model is obtained by solving the evolution equation incorporating the known physics effects instead of simple extrapolation. Therefore, such a semi-analytical approach is highly complementary to the simulation efforts and the two approaches can lead to insights to help improve each other. We significantly updated and expanded the related studies initiated in our earlier paper \cite{Chang:2019mza}, taking into consideration the very recent developments since then. For instance, \cite{Blasi:2020wpy,Cui:2019kkd} show that the inclusion of the very high oscillation modes can drastically change the shape of the GW spectrum from NG strings in (early-)matter dominated era, which was neglected in earlier literature. We included the contribution from these high modes in this updated study, which leads to substantial modifications to the GW spectrum at low $f$ for standard thermal history as well as at high $f$ with the presence of an early matter domination epoch. We also discuss the consequence for the prediction of SGWB if the non-scaling behavior found in some simulation results for early evolution sustains in the late-time evolution of a global string network, compare with the results found in \cite{Gorghetto:2021fsn,Klaer:2019fxc}, and suggest potential modifications to the VOS model to accommodate such a feature. We will dive into the time-frequency correspondence for global strings, which is the guiding principle for testing standard cosmology. We conduct an extensive study on probing a potentially existing non-standard equation of state of the pre-BBN Universe such as early matter domination (EMD) or kination, where we also include a concrete example for a finite duration of a kination epoch. In addition, we study the effects on the GW spectrum with the presence of new massive degrees of freedom. Furthermore, a detailed discussion is given to address uncertainties such as loop size distribution, radiation parameters, and distinguishing from other SGWB sources. Our results directly apply to pure global strings associated with massless Goldstone. The application to the axion case where the Goldstone acquires a mass at a QCD(-like) phase transition is more complex and requires treatments of the axion domain walls in addition to the strings, see for example \cite{Gelmini:2021yzu}. We reserve a  dedicated study on the axion case for future work. We also comment on the prospect of addressing the recent NANOGrav result with global strings. 

The rest of this article is organized as follows. In Section \ref{sec:VOS} we will present our methodology based on the analytical Velocity-dependent One-Scale (VOS) model for global strings calibrated with recent simulation results. In Section \ref{sec:DoGSL} we derive the GW frequency spectrum from a global string network in the context of standard thermal history. In Section \ref{sec:GWSGCS} we illustrate the relation between the frequency of a GW signal observed today and its emission time in the early Universe. With several benchmark examples, we show how this relation can be used to test standard cosmology and detect potential new physics. Related experimental constraints and sensitivities are also demonstrated in Section \ref{sec:GWSGCS} and \ref{sec:DEUGS}. In Section \ref{sec:Diss} we will address various uncertainty factors that may affect the results, as well as how to distinguish global string induced SGWB from other potential SGWB sources. We make our conclusions in Section \ref{sec:con}.

\section{Evolution of a Global Cosmic String Network}

\subsection{Velocity-dependent One-Scale (VOS) model for global strings}
\label{sec:VOS}
Recent years have seen rapid developments in simulating a global/axion string network \cite{Gorghetto:2020qws,Gorghetto:2018myk,Fleury:2015aca,Saurabh:2020pqe,Hindmarsh:2021vih,Hindmarsh:2019csc,Martins:2020jbq,Klaer:2019fxc,Buschmann:2019icd,Vaquero:2018tib}. Nevertheless, a technical challenge persists for pure numerical simulation to track the network's evolution over the entire relevant cosmic history. Two characteristic scales need to both be captured by simulation: the string width which is about the inverse of the related symmetry breaking scale $r_{\hbox{\st{core}}} \sim 1/\eta$, the time-dependent horizon size of the Universe which is of the Hubble scale $H^{-1}$. There is generally a large hierarchy between the two scales, which can be up to $\eta/H \sim 10^{57}$ in the late-time universe. However, current simulations can only cover very early stage of the evolution up to $\eta/H \sim 10^{3}$, therefore extrapolation, potentially unreliable for late times, has to be made to make prediction for observations today. Our approach here is to adopt an analytical VOS model that captures the essential physics, and use it to study and predict the evolution of the string network over a long range of time, while calibrating the input model parameters with data points for early time evolution that have been made available by simulations. 

In this section, we review the VOS model of a global string network and compare its predictions with that from simulations. The VOS model was originally introduced in the context of NG strings \cite{Martins:1996jp,Martins:2000cs,Martins:2003vd}, and recently extended/updated including the application to axion strings \cite{Correia:2019bdl,Martins:2016wqq,Martins:2018dqg}. The VOS model has been widely supported by simulation results in the case of NG strings \cite{Blanco-Pillado:2013qja,Blanco-Pillado:2017oxo,Blanco-Pillado:2017rnf}, yet for global strings it is still being tested by simulations. According to the VOS model, starting with an arbitrary initial condition, the cosmic string network would eventually enter a scaling regime \cite{Klaer:2019fxc,Gorghetto:2018myk}, where the correlation length $L$ (or the mean of the inter-string separation scale) of the strings remains constant relative to the horizon size, and the energy density of the network tracks the total background energy density with a coefficient $\sim G\mu$. The network typically consists of a few horizons sized long strings along with copious sub-horizon sized string loops. In this regime, the energy density of the string network relative to the background energy density does not grow with the scale factor $a$ due to the energy loss from the decay of the loops. While GWs constitute the leading radiation by the NG strings, they are irreducible but subdominant mode for global strings for which the emission of Goldstone particles is more important\footnote{We neglect the emission of radial mode which is shown to decouple soon after the network formation \cite{Gorghetto:2021fsn,Gorghetto:2020qws} and may be generally suppressed when the loop size is larger than $\sim 1/\eta$ \cite{Saurabh:2020pqe}.}. The energy density of the global string network (mainly stored in long strings) is
\begin{align}
\label{Eq: rho}
\rho_{\infty} = \frac{\mu(t)}{L^2(t)} = \xi(t) \frac{\mu(t)}{t^2},
\end{align}
where the dimensionless parameter $\xi(t)$ is defined as the number of long strings per horizon volume. $\mu(t)$ is the time-dependent tension (i.e. energy per unit length) of the global strings ($\mu$ is a constant for NG or local strings),
\begin{align}
\mu(t) = 2 \pi \eta^2 \hbox{ln}\frac{L}{\delta}\equiv2\pi\eta^2N, 
\end{align}
with
\begin{align}
\delta \sim (\sqrt{\lambda} m_\phi)^{-1} \;\;\;\; \hbox{and} \;\;\;\; m_\phi^2 = \lambda \left|\left(\frac{T^2}{3} - \eta^2 \right)\right| \sim \lambda \eta^2,
\end{align}
where $\delta$ is the width of the string core, $\lambda$ is the coupling in $\phi^4$ theory and $m_\phi$ sets the mass of the Higgs-like complex  $\phi$ whose VEV breaks the global $U(1)$, and we have defined the time-dependent parameter $N$ which will be used in later discussions. The temperature $T$ dependent thermal mass contribution is negligible well after the symmetry breaking phase transition ($T\ll \eta$), and thus we ignore it in our analysis. We consider $\lambda \sim 1$ such that $m_\phi$ and $\eta$ are comparable. The evolution equation for the correlation length $L$ is \cite{Vilenkin:2000jqa,Martins:2018dqg,Martins:1996jp,Martins:2000cs}
\begin{align}
\label{Eq2-3}
\left(2 - \frac{1}{N} \right) \frac{dL}{dt} = 2 H L \left( 1+ \bar{v}_\infty^2 \right) + \frac{L \bar{v}_\infty^2}{\ell_f} + \bar{c}\bar{v}_\infty  + s \frac{\bar{v}_\infty^6}{N},
\end{align}
which couples to the evolution equation for the average long string velocity $\bar{v}_\infty$:
\begin{align}
\frac{d\bar{v}_\infty}{dt} = \left( 1-\bar{v}_\infty^2 \right) \left[ \frac{k_v}{L}- 2 H\bar{v}_\infty \right],
\end{align}
where $k_v$ is the momentum parameter. While we will investigate the detectability of GW signal, we left out the GW radiation term in these evolution equations because its contribution here is sufficiently suppressed \cite{Vilenkin:2000jqa}. The terms on the RHS of Eq.~\ref{Eq2-3} represent, in order, the dilution effect from the expansion of the Universe, thermal friction effect with characteristic scale $\ell_f \propto \mu T^{-3}$, loop chopping rate parameter $\bar{c}$, and the back-reaction due to Goldstone boson emission \cite{Martins:2018dqg,Martins:2000cs}. The thermal friction is negligible as the Universe cools down such that $T\ll \eta$. \\

In the following analysis we consider various possibilities of background cosmology parametrized by $n$, defined as
\begin{align}
\label{Eq2-4}
\rho \propto a^{-n}, \;\;\;\;\;a(t) \propto t^{2/n}
\end{align}
where $a(t)$ is the expansion parameter as a function of time $t$, $\rho$ is the background cosmic energy density. $n=3,4$ correspond to the cases of matter (MD) and radiation (RD) domination, respectively. We focus on the range of $2 < n \leq 6$ ($n=6$ corresponds to kination epoch which we will discuss more in Sec.~\ref{sec:DEUGS}). 

In the scaling regime, the parameters $\xi$ and $\bar{v}_\infty$ are approximately time-independent. For a specific $n$, the solution to the evolution equations in the VOS model can be expressed as
 \cite{Martins:2018dqg}
\begin{align}
\label{Eq2-5}
\xi= \left( \frac{L}{t} \right)^{-2}=\frac{8 \left( 1 - \frac{2}{n} - \frac{1}{2N} \right)}{nk_v(k_v+\bar{c})(1+\Delta)}, \;\;\;\;\; \bar{v}_\infty^2 \equiv v_0^2(1-\Delta)= \frac{n-2 - \frac{n}{2N}}{2} \frac{k_v}{k_v+\bar{c}}\left(1-\Delta\right),
\end{align}
with
\begin{align}
\label{Eq2-6}
\Delta \equiv \frac{\sigma}{N (k_v+\bar{c})}, \;\;\;\;\;\;\; \sigma\equiv sv_0^5.
\end{align}
The Goldstone particle radiation term $s \bar{v}_\infty^6/N$ is treated as a perturbation (valid when $\Delta \ll 1$), and $v_0$ is the solution to $\bar{v}_\infty$ in the limit where the Goldstone emission term is set to 0. The model parameters $\{\bar{c}, k_v, \sigma\}$ can be extracted by calibrating with current simulation results, as we will discuss. \\

Although the presence of a scaling regime with a constant $\xi$ as in the VOS model has been confirmed by simulations for NG or gauge strings \cite{Hindmarsh:2011qj,Albrecht:1997mz,Pogosian:1999np,Avgoustidis:2012gb,Hindmarsh:2017qff,Hindmarsh:2018wkp}, the situation is not yet clear for global strings. Some of the recent simulation studies such as \cite{Gorghetto:2020qws} suggest a time-dependent $\xi(t)$ that grows linearly with $N$ in small $4 \lesssim N \lesssim 7$ region, 
\begin{align}
\label{Eq2-6-2}
\xi(t) \simeq 0.24(2) \times N + \beta,
\end{align}
where $\beta$ is a constant bearing large uncertainty related to initial condition. The linear increase of $\xi$ in $N$ is also found in some other simulation studies \cite{Vaquero:2018tib,Klaer:2019fxc,Klaer:2017qhr,Kawasaki:2018bzv,Buschmann:2019icd,Fleury:2015aca}, but is in conflict with other groups' simulation results, which predict a nearly constant $\xi$ \cite{Figueroa:2020lvo,Hindmarsh:2019csc,Hindmarsh:2021vih,Hindmarsh:2017qff,Martins:2003vd,Lopez-Eiguren:2017,Yamaguchi:2002sh,Hiramatsu:2010yu,Hiramatsu:2012gg,Kawasaki:2014sqa}. This discrepancy is an intriguing puzzle, and requires further investigation with higher resolution simulations. Given the uncertainty, while we mainly focus on the application of the VOS model, here and in Sec.~\ref{Sec:Non-scaling solution} we also carefully considered the effect of potential deviation from scaling and suggest modification/extension to the current VOS model. 

In order to calibrate the parameters $\{\bar{c}, k_v, \sigma\}$ for the VOS model, we fit data extracted from simulation results in \cite{Klaer:2017qhr,Gorghetto:2018myk,Hindmarsh:2019csc}, as summarized in Table.~\ref{Table_Data}. The error bars are visually estimated from the plots in \cite{Klaer:2017qhr,Gorghetto:2018myk,Hindmarsh:2019csc}, as we are doing a simplified statistical analysis as in \cite{Martins:2018dqg}.
Our best fitting result for the VOS model parameters are as follows:
\begin{align}
\label{Eq: Fitting Result}
\{\bar{c}, k_v, \sigma\} \simeq \{0.497,0.284,5.827\},
\end{align}
and the fitting quality is about $3.3$-$\sigma$ significance (p-value $<$ 0.001). Such a fitting quality reflects moderate tensions among simulation data listed in Table.~\ref{Table_Data}, possibly due to the different simulation methods as well as the different ways of counting the number of strings that are employed in the literature. We will assume that these same parameters apply for different scenarios of cosmological background, e.g. radiation domination (RD) or matter domination (MD) \footnote{While this assumption has been confirmed for NG/local strings, a recent simulation work for global strings \cite{Klaer:2019fxc} suggests that $\bar{c}$ may differ with different background cosmologies. The deviation mostly originates from the difference in the predicted averaged velocity $\bar{v}_\infty$, while $\xi$ is about the same in different cases.}. 
As an example, for $N=70$, we obtain the number of strings per Hubble volume $\xi\sim 4.0$ and $\bar{v}_\infty \sim 0.57$ in RD, and $\xi \sim 3.55$ with $\bar{v}_\infty \sim 0.40$ in MD. 

\begin{figure}[t]
\centering
\includegraphics[width=0.9\textwidth]{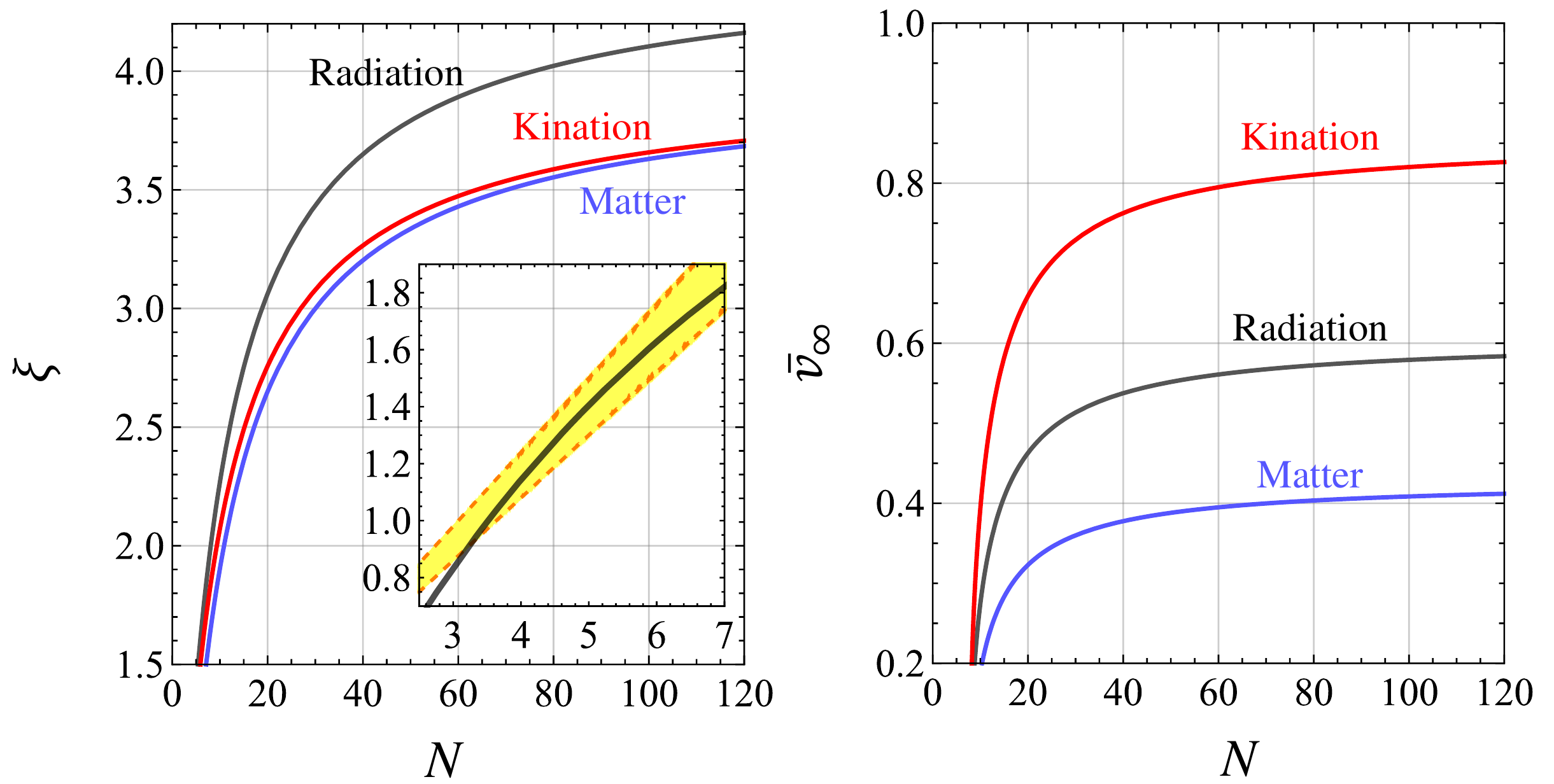} 
\caption{\label{F1} The number of global string per Hubble volume $\xi$ and the average long string velocity $\bar{v}_\infty$ as functions of $N \equiv \hbox{ln}\frac{L}{\delta}$, as predicted by the VOS model, for different background cosmologies. The subfigure in the left panel is the zoom-in of $\xi$ evolution in the low $N$ range during radiation domination, where the yellow band shows the 1-$\sigma$ uncertainty region based on the finding by simulation (as shown in Eq.(\ref{Eq2-6-2}) with $\beta \sim 0.20$).}
\end{figure}

In Fig.~\ref{F1} we show the evolution of $\xi$ and $\bar{v}_\infty$ as functions of $N$ using the VOS model with the fitting model parameters listed above. Given the recent findings suggesting deviation from scaling (Eq.(\ref{Eq2-6-2})), in the sub-figure of Fig.~\ref{F1} where the small $N$ region is zoomed in, we also show the $1$-$\sigma$ area of Eq.(\ref{Eq2-6-2}) as the yellow band (the error bar is given in \cite{Gorghetto:2020qws}), in comparison with the VOS model prediction (sold curve). We found that in the region of small $N \lesssim 7$ the VOS model prediction is consistent with a linear growth of $\xi$ in $N$, provided that $\beta$ is not too small, e.g. $\beta \sim 0.20$ is taken as an example in our analysis. The late-time evolution in the scaling regime is insensitive to the exact value of $\beta$ which depends on initial conditions. Nevertheless, as can be seen in Fig.~\ref{F1}, at large $N$ VOS model prediction approaches scaling, i.e. a nearly constant $\xi$. For our later analysis of the GW signals, the late-time evolution in the region of $N \gtrsim 50$ region is most relevant, yet is beyond the reach of most of current simulations. In contrast, VOS model provides a reasonable prediction for the entire time range of interest, after calibrating with low $N$ simulation data.

The data points we used and listed in Table.\ref{Table_Data} were also applied in \cite{Martins:2018dqg}. We initially considered using a larger data set for the fitting by including more simulation results, but they are in some way in conflict with the data in Table.\ref{Table_Data}. In order to have meaningful results, we decided to leave out the data sets that fit VOS model poorly. The discrepancies among different simulation results could be in part because these simulations are done with very different methods, covering different ranges of $N$, and the number of strings and the velocities are counted by different numerical algorithms \cite{Martins:2018dqg}. Further investigations and developments are certainly required to reach a convergence among different simulation results. To fairly consider these other data sets, in the following, we further discuss their implications and why we left them out of our analysis.

First, note that the VOS model is only valid once the string network enters the scaling region $N \sim 6$ (e.g. see Fig.~3 of \cite{Gorghetto:2018myk}). The evolution in the very early stage of $N \lesssim 5$ is sensitive to initial condition. Therefore, for our fitting, we exclude simulation data points with very low $N$'s such as in \cite{Kawasaki:2018bzv} ($N=2-4$). The result from \cite{Hindmarsh:2021vih} is not included because we found that its large velocity $\bar{v}=0.609\pm 0.014$ leads to a poor $\chi^2$ fit with other simulation data\footnote{Alternatively, one could include the result from \cite{Hindmarsh:2021vih} into the $\chi^2$ analysis. We found that in this case, the fitting gives $\{\bar{c}, k_v, \sigma\} \simeq \{ 0.588,0.395,0.314 \}$ with $p$ value $\ll 0.01\% $, which indicates a poor fit. Based on this result, we found that $\xi$ parameter would be decreased by a factor of 2, without significant change to $\bar{v}_\infty$ in the large $N$ range}. Consequently, with these parameters the GW production would be reduced to about 44\% compare to the GWs prediction with parameter set Eq.(\ref{Eq: Fitting Result}) as we will see in later section. Ref.~\cite{Klaer:2019fxc} simulated the global string network with cosmological background parameter $n \leq 3$, without a data point simulated with a radiation dominated background, thus cannot be analyzed with the results included in Table.~\ref{Table_Data}. Another reason we did not include data from \cite{Klaer:2019fxc} is that their results suggest a time-dependent loop chopping rate, which does not match the VOS model. \cite{Buschmann:2019icd}
suggests another pattern of deviation from scaling that is inconsistent with Eq.(\ref{Eq2-6-2}). While these suggested non-scaling behaviors only directly apply to low $N$ range and do not converge among literature, it is intriguing to consider their potential effects on GW signals (if the non-scaling persist till large $N$) and how VOS model would need to be revised accordingly. We leave more discussion on this topic in Sec.~\ref{Sec:Non-scaling solution}. 
  
In this study, we simply keep the velocity parameter $k_v$ as a constant as in the conventional VOS model. Nevertheless, some studies suggested the possibility of velocity-dependent momentum parameter $k_v = k_v(v)$ \cite{Martins:2003vd, Martins:2016ois,Correia:2019bdl,Correia:2020gkj} \begin{align}
\label{Eq: kv}
k_v(v) = k_0 \frac{1-(q v^2)^\beta}{1+(q v^2)^\beta},
\end{align}
where $q\simeq 2.3$, $\beta \simeq 1.5$, and $k_0\simeq 1.37$ \cite{Correia:2019bdl,Correia:2020gkj}. We found that in RD background Eq.(\ref{Eq: kv}) gives a numerical value of velocity parameter $k_v(v=\bar{v}_\infty) \sim 0.3$ at high $N \gtrsim 10$ which is consistent with our fitting result Eq.(\ref{Eq: Fitting Result}). In addition, there is a debate about whether the chopping parameter $\bar{c}$ is time-independent: e.g.  \cite{Klaer:2019fxc} suggests that $\bar{c}$ decreases with $N$, while \cite{Hindmarsh:2021vih} fits a constant value $\bar{c}=0.843\pm 0.039$ in radiation background. We will not elaborate on these two particular types of uncertainty. 
\begin{table}[t]
\centering
\begin{tabular}{l| ccc}
  \hline
    \hline
  \hbox{Reference} & $N$ & $\xi$ & $\bar{v}$ \\
  \hline
\hbox{Klaer et al.} \cite{Klaer:2017qhr} &$ 55$ &$ 4.4 \pm 0.4$ & $0.50 \pm 0.04$ \\
 &$ 31$ &$ 4.0 \pm 0.4$ & $0.50 \pm 0.04$ \\
  &$ 15$ &$ 2.9 \pm 0.3$ & $0.51 \pm 0.04$ \\
 \hbox{Gorghetto et al.} \cite{Gorghetto:2018myk} &$ 6-7$ &$ 1.0 \pm 0.30$ &  \\
 \hbox{Hindmarsh et al.} \cite{Hindmarsh:2019csc}&$ 6$ &$ 1.19 \pm 0.20$ &  \\ 
    \hline
      \hline
\end{tabular}
\caption{Results from recent global string network simulations (in a radiation dominated background) for the number of strings per Hubble volume $\xi$ and the average velocity of long strings $\bar{v}$ in radiation dominated background. These data points were also applied in \cite{Martins:2018dqg}. In the main text, we explain why some other recent simulation results were left out of this table (thus our analysis) and their implications. \label{Table_Data}}
\end{table}%

\subsection{Dynamics of global string loops: formation and radiation into GWs and Goldstones}\label{sec:DoGSL}

A global cosmic string network forms during the phase transition around $T \sim \eta$. The dynamics of the very early stage of evolution is sensitive to initial conditions. However, the string network would soon evolve towards an initial condition independent scaling regime \cite{Vaquero:2018tib,Gorghetto:2020qws,Gorghetto:2018myk,Klaer:2019fxc,Klaer:2017qhr}, namely, $\xi \sim \,$constant (or with potential deviation from scaling suggested by some recent work, see earlier discussion and later in Sec.~\ref{Sec:Non-scaling solution} ). The horizon-sized long strings randomly intersect each other and lose energy via forming sub-horizon sized loops, which subsequently oscillate and radiate energy until they decay away. The loop size distribution at formation time can be parameterized by: a distribution function $\mathcal{F}_\alpha$ and the fraction of energy stored in loops that can be released as radiation (GWs or Goldstones), $F_\alpha$. In this work, we consider two representative scenarios in detail, both inspired by simulation results: (1) a nearly monochromatic loop size at formation $\ell_i\sim\alpha t_i$ with $\alpha \sim 0.1$, such that $\mathcal{F}_{\alpha\sim 0.1} \sim 1$, while $\sim90\%$ fraction of loop's energy is in the form of kinetic energy which would eventually redshift away without contributing to GWs, thus $F_\alpha \sim 0.1$, as inspired by  \cite{Blanco-Pillado:2017oxo,Blanco-Pillado:2013qja}; and (2) a flatter, log-uniform distribution of loop size as suggested in \cite{Gorghetto:2018myk}. In this section we focus on the simpler first case, and the second scenario will be discussed in Sec.~\ref{Sec:4-5}. In Sec.~\ref{Sec:4-5} we also comment on other loop distribution possibilities to account for the related uncertainties \cite{Auclair:2019wcv}. 
 
 \begin{figure}[t]
\centering
\includegraphics[width=0.48\textwidth]{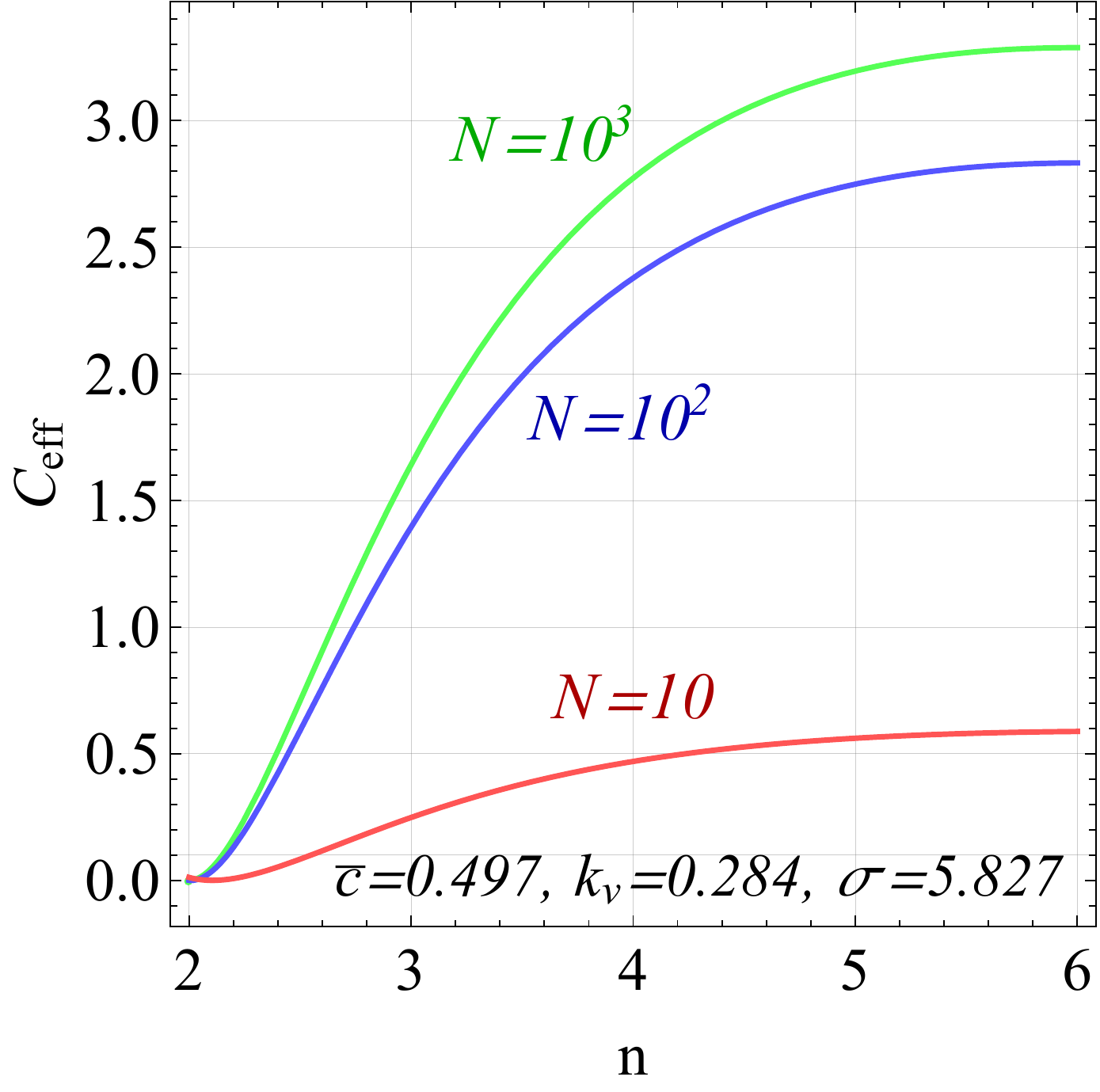} 
\includegraphics[width=0.48\textwidth]{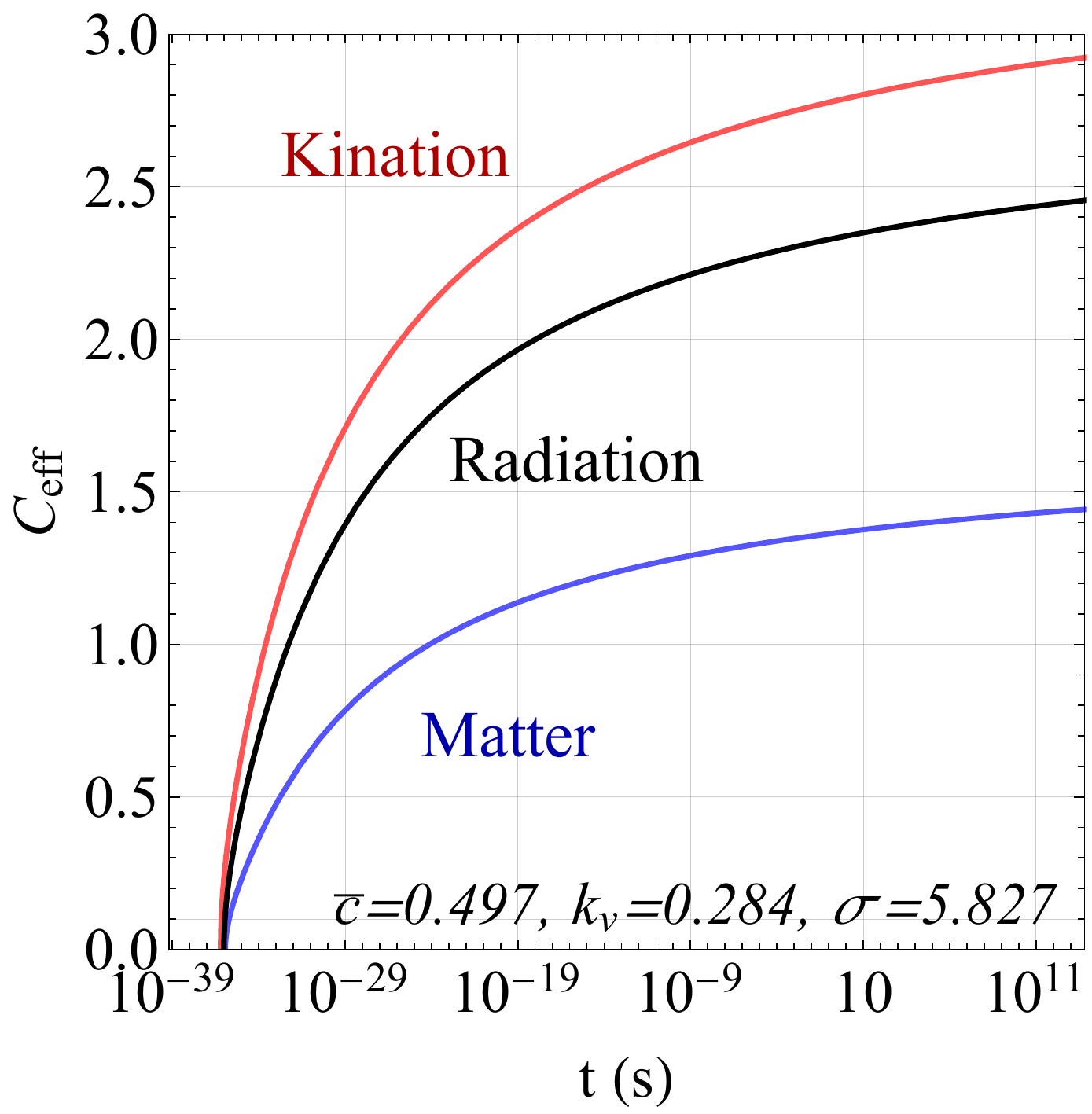} 
\caption{\label{F2} The dependence of loop emission factor $C_{\hbox{\st{eff}}}$ on the background cosmology as derived from the VOS model, as well as its evolution in time (characterized by $N\equiv \hbox{ln}\frac{L(t)}{\delta}$ or $t$). In the example shown, the symmetry breaking scale is taken as $\eta=10^{15}\,$GeV.}
\end{figure}

By energy conservation, for a specific $\alpha$ the formation rate of string loops in a scaling string network is given by 
\begin{align}
\label{Eq2-7}
\frac{d \rho_o}{dt} \times \mathcal{F}_\alpha = -\left( \frac{d\rho_\infty}{dt} \right) \times F_\alpha \times \mathcal{F}_\alpha = \bar{c} \bar{v}_\infty \frac{\mu}{L^3} F_\alpha \mathcal{F}_\alpha,
\end{align}
where the chopping rate parameter $\bar{c}$ is given in Eq.(\ref{Eq: Fitting Result}), and $\rho_o$ denotes the energy density of string loops. The number density of loops with length $\ell = \alpha t$ is then
\begin{align}
\label{Eq2-8}
dn_o = \frac{\bar{c}\bar{v}_\infty \mathcal{F}_\alpha F_\alpha}{\alpha t L^3} dt \equiv \mathcal{F}_a C_{\hbox{\st{eff}}} \frac{F_\alpha}{\alpha}\frac{dt}{t^4},
\end{align}
where we define the loop emission parameter $C_{\hbox{\st{eff}}}$ 
\begin{align}
\label{Eq: Ceff}
C_{\hbox{\st{eff}}} \equiv \bar{c} \bar{v}_\infty \xi^{3/2}.
\end{align}
We obtain $C_{\hbox{\st{eff}}}$ for different background cosmologies (i.e. equations of state) based on the solutions given in Eq.(\ref{Eq: Fitting Result}) and Eq.(\ref{Eq2-5}). 
Fig.~\ref{F2} illustrates the solution and evolution of $C_{\hbox{\st{eff}}}$. Numerically, we found $C_{\hbox{\st{eff}}}\simeq \{ 1.32, 2.26, 2.62, 2.70 \}$ for $n=\{3,4,5,6\}$ ($n$ parameterizes cosmology as defined in Eq.(\ref{Eq2-4}), respectively.
Note that $C_{\hbox{\st{eff}}}$ falls down to zero when the $\bar{v}_\infty \to 0$, which corresponds to the time $t_\eta$ when the Goldstone radiation becomes important in the equation of motion Eq.(\ref{Eq2-3}), i.e. $\Delta(t=t_\eta) = 1$ in the string network evolution (see Eq.(\ref{Eq2-5})):
\begin{align}
\label{Eq2-10-2}
t_\eta \simeq \frac{\sqrt{\xi(t_\eta)}}{\eta} \exp\left( \frac{\sigma}{k + \bar{c}} \right).
\end{align}
The right panel of Fig.~\ref{F2} illustrates this point with numerical results. With our calibrated parameters, $t_\eta$ as defined corresponds to  $N\sim 6-7$, which implies that the perturbative VOS model \cite{Martins:2018dqg} has large uncertainties in such a low $N$ range. 

After formation, a global string loop would rapidly oscillate and emit energy in the form of GWs and Goldstones by the following energy loss rates until the loop disappears completely \cite{Vilenkin:1986ku}:
\begin{align}
\label{Eq3-1}
\frac{dE}{dt} = -\Gamma G \mu^2 - \Gamma_a \eta^2.
\end{align} 
 Note that the parameter $\Gamma_{(a)}$ only depends on the loop trajectory \cite{Vilenkin:2000jqa,Battye:1997jk}, thus we expect that the Goldstone radiation constant $\Gamma_a$ should be close to the value of the GW radiation constant $\Gamma \sim \Gamma_a$ \cite{Battye:1997jk} which is also determined by the loop shape. In the following, we assume benchmark values $\Gamma \sim 50$ \cite{Vilenkin:1981bx,BlancoPillado:2011dq,Blanco-Pillado:2013qja,Blanco-Pillado:2017oxo}, and $\Gamma_a \sim 65$ \cite{Battye:1997jk,Vilenkin:2000jqa}. We will discuss the effect of varying $\Gamma_a$, $\Gamma$ in Sec.~\ref{sec:Gamma} to account for the potential uncertainty on the radiation parameters.

The size of a loop with initial length $\ell_i = \alpha t_i$ therefore decreases as
\begin{align}
\label{Eq3-2}
\ell(t) \simeq \alpha t_i - \Gamma G \mu (t-t_i) - \frac{\Gamma_a}{2\pi} \frac{1}{\hbox{ln}(L/\delta)}(t-t_i),
\end{align}
where $t_i$ is the loop formation time. The radiation of GW and Goldstone from a loop can be decomposed into a set of normal-mode oscillations with frequencies $\tilde{f}_k = 2 k /\tilde{\ell}$, where mode numbers $k=1,2,3 \cdots$, and $\tilde{\ell} \equiv \ell(\tilde{t})$ is the instantaneous size of the loop when it radiates at $\tilde{t}$. We can rewrite the radiation parameters in a decomposed form
\begin{align}
\label{Eq: Gamma k modes}
\Gamma^{(k)} = \frac{\Gamma k^{-\frac{4}{3}}}{\sum_{m=1}^\infty m^{-\frac{4}{3}}},
\;\;\;\;\; \hbox{and} \;\;\;\;\; \Gamma^{(k)}_{a} = \frac{\Gamma_a k^{-\frac{4}{3}}}{\sum_{m=1}^\infty m^{-\frac{4}{3}}},
\end{align}
where $\sum_{m=1}^\infty m^{-\frac{4}{3}} \simeq 3.60$, $\sum_k \Gamma^{(k)} = \Gamma$, and $\sum_k \Gamma^{(k)}_{a} = \Gamma_{a}$. We have assumed that the cusps are the dominating source of GW and Goldstone emissions as found in NG string simulations \cite{Olum:1998ag,Blanco-Pillado:2015ana,Blanco-Pillado:2019nto}. The contributions from kinks and kink-kink collisions follow different power laws: $\Gamma^{(k)} \propto k^{-5/3}$ and $k^{-2}$ for kinks and kink-kink collisions, respectively \cite{Vachaspati:1984gt,Burden:1985md,Garfinkle:1987yw}. As shown in Eq.(\ref{Eq3-1}), relative to Goldstone emission, GW radiation is suppressed by a factor of $\sim \eta^2/m_p^2$, where $m_p$ is Planck scale. 
Nevertheless, the suppression factor becomes less severe as the symmetry breaking scale $\eta$ gets closer to $m_p$.

Our main analysis results shown in Sec.~\ref{sec:GWSGCS} are obtained by focusing on the simple, motivated assumptions made in this section. Nevertheless, we acknowledge other possibilities of $C_{\hbox{\st{eff}}}$ and $\Gamma_{(a)}$ that were suggested in literature. We further discuss the effects on phenomenology in light of possible deviations from our assumptions on these factors in Sec.~\ref{sec:Gamma} and Sec.~\ref{Sec:Non-scaling solution}.  
\section{SGWB Spectrum from Global Cosmic Strings (Standard Cosmology)}\label{sec:GWSGCS}
 In this section we will first show the derivation and numerical results of SGWB frequency spectrum from a global cosmic string network assuming a standard cosmic history (Sec.~\ref{sec:SGSN}, \ref{sec:GWFSaES}). Then in order to give more physics explanation and insights, in Sec.~\ref{sec:RDGG} we provide parametric estimates for the relic densities of Goldstones and GWs emitted from global strings, and compare with GW signals from NG strings. Comparison with related results in other literature is also given.

\subsection{Derivation of GW spectrum from global strings}
\label{sec:SGSN}
The generic form of the relic energy density of a SGWB is given by
\begin{align}\label{Eq: f ell}
\Omega_{\hbox{\st{GW}}} = \frac{f}{\rho_c} \frac{d\rho_{\hbox{\st{GW}}}}{df},
\end{align}
where $\rho_{\hbox{\st{GW}}}$ is the energy density of GWs, and $\rho_c = 3 H_0^2/8\pi G$ is the critical density. 
String loops emit GWs from normal mode oscillations with frequencies $\tilde{f}_k=\frac{2 k}{\tilde{\ell}}$, where $k\in\mathbb{Z}^+$, $\tilde{\ell}$ is the loop size at emission time $\tilde{t}$ \cite{Chang:1998tb,Gorghetto:2020qws}. Taking into account of redshift effects, the observed frequencies today are then
\begin{align}
\label{Eq3-5-2}
f_k = \frac{a(\tilde{t})}{a(t_0)} \tilde{f}_k = \frac{2 k}{\tilde{\ell}}  \frac{a(\tilde{t})}{a(t_0)}.
\end{align}
  The relic GW background is obtained by summing over all harmonic modes
\begin{align}
\Omega_{\hbox{\st{GW}}}(f) = \sum_k \Omega_{\hbox{\st{GW}}}^{(k)}(f) = \sum_k  \frac{f_k}{\rho_c} \frac{d\rho_{\hbox{\st{GW}}}}{df_k}.
\end{align}

Using Eq.(\ref{Eq2-8}) and Eq.(\ref{Eq3-2}) that we derived earlier and integrating over emission time $\tilde{t}$, we can derive the contribution $\Omega_{\hbox{\st{GW}}}^{(k)}(f)$ from an individual $k$ mode as
\begin{align}
\label{Eq3-6}
\Omega_{\hbox{\st{GW}}}^{(k)}(f) = \frac{\mathcal{F}_\alpha}{\rho_c} \frac{2 k}{f} \frac{F_\alpha}{\alpha} \int_{t_F}^{t_0} d\tilde{t} \frac{\Gamma^{(k)} G \mu^2}{\left( \alpha + \Gamma G \mu + \frac{\Gamma_a}{2\pi N}\right)} \frac{C_{\hbox{\st{eff}}}\left(t_i^{(k)} \right)}{t_i^{(k)4}} \Theta(t_i,\tilde{t}) \left( \frac{a(\tilde{t})}{a(t_0)}\right)^{5} \left( \frac{a(t_i^{(k)})}{a(\tilde{t})} \right)^{3}\,
\end{align}
where $t_F$ is the formation time of the global string network, $t_0$ is the current time, and the causality and energy conversation conditions are imposed by
\begin{align}
\label{Eq3-7}
\Theta(t_i,\tilde{t}) = \theta(\tilde{\ell})\, \theta(\tilde{t} - t_i).
\end{align}
With Eq.(\ref{Eq3-2}) we can derive that a loop that emits GW at time $\tilde{t}$ leading to an observed frequency $f$ was formed at the time
\begin{align}
\label{Eq:ti k}
t_i^{(k)} (\tilde{t},f) = \left( \frac{1}{\alpha + \Gamma G \mu + \frac{\Gamma_a}{2\pi N}} \right) \left[ \tilde{\ell}(\tilde{t},f,k) + \Gamma G \mu \tilde{t} + \frac{\Gamma_a}{2\pi N} \tilde{t} \right],
\end{align}
Note that to consider the radiation of Goldstones, we may define $\Omega_{\hbox{\st{Gold}}}(f)$ in analogy to Eq.(\ref{Eq3-6}) with the simple replacements: $\Gamma \to \Gamma_a,$ and $\Gamma G \mu^2 \to \Gamma_a \eta^2$. We will apply this prescription in later discussions involving Goldstone radiation (e.g. Sec.~\ref{sec:RDGG}).

\begin{figure}[t]
\centering
\includegraphics[width=0.94\textwidth]{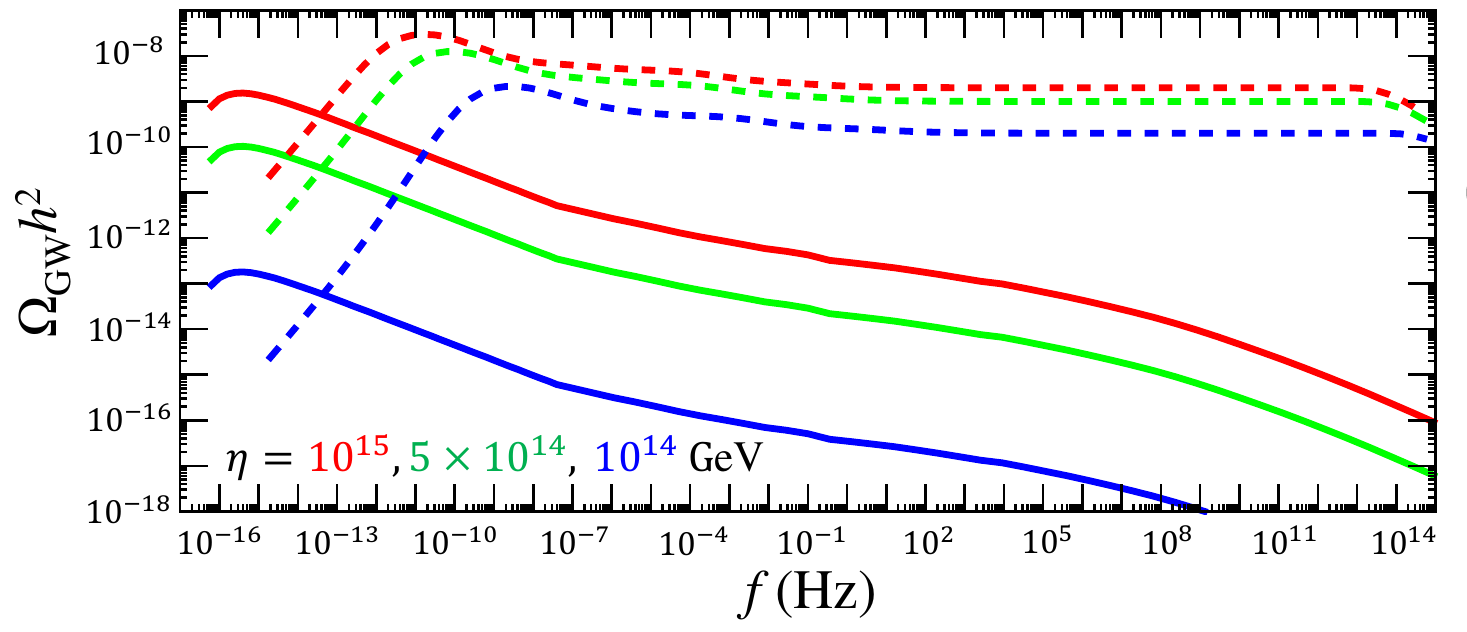} 
\caption{\label{F3} Gravitational wave spectra from a global (solid) and NG (dashed) string network with $\alpha = 0.1$, $F_\alpha=0.1$, for  $\eta = 5\times 10^{15}~({\rm red}), \, 10^{15}~({\rm orange}), 5\times 10^{14}~({\rm green}), and 10^{14}$ GeV ~({\rm blue}). Up to $k=10^{15}$ harmonic modes are included in the summation.}
\end{figure}

Earlier studies based on radiation dominated background \cite{Battye:1997jk,Davis:1989gn,Davis:1989nr} found that the first few $k$-modes dominate the GW radiation from loops. However, recent work (in the context of NG strings) showed that a large value of $k\gtrsim10^5$ may be needed to converge, depending on the background cosmology \cite{Blasi:2020wpy,Cui:2019kkd,Gouttenoire:2019kij}. For instance, including higher $k$ modes changes the power-law index of $\Omega_{\hbox{\st{GW}}}(f)$ from $-1$ to $-1/3$ in a MD epoch. In this work we investigated the importance of high $k$ modes in the context of global strings and found similar results. We found that to reach a converging result for $\Omega_{\hbox{\st{GW}}}(f)$ up to $\sim100$ Hz, i.e. within the frequency range relevant for current and near-future GW detections, up to $k\sim10^8$ modes need to be included. Higher $f$ range requires more $k$ modes to converge. To draw the full spectrum shown in Fig.~\ref{F3} we took into account of up to $k\sim10^{15}$ modes in our analysis. 

Fig.~\ref{F3} demonstrates the SGWB spectrum calculated numerically with the method we outlined. The corresponding results for NG strings are also shown in comparison, with more explanation given in Sec.~\ref{sec:RDGG}. While the very high $f$ range $f\gg100$ Hz is well beyond the reach of any foreseeable experiment, we keep it in Fig.~\ref{F3} for theoretical completeness by capturing physics at times as early as the formation time of the string network. As can be seen, towards high $f$ the spectrum falls most significantly starting around the frequency $f_\eta \sim \frac{2}{\alpha t_\eta} [a(t_\eta)/a(t_0)] \sim 10^{10}\,$Hz \cite{Battye:1996pr}, as a result of the string network formation time and the validity cutoff of the perturbative VOS model (Eq.(\ref{Eq2-10-2})). The exact shape of the falling spectrum at frequencies $f>f_\eta$ has uncertainties and is sensitive to the initial condition and the very early stage of string network evolution, which is not captured by the VOS model. Then over a wide range of $f$ the spectrum gradually declines towards higher $f$ ($\sim\ln^3 (1/f)$, see Eq.(\ref{Eq5-1}) below), corresponding to the emissions during the RD era. Note that this feature of the SGWB spectrum from global strings is in contrast to a nearly flat plateau as in its NG string counterpart. Starting at $f_0 \simeq \frac{2}{\alpha t_0} \sim 3.6\times 10^{-16}\,$Hz, the spectrum behaves as $f^{-1/3}$ until $f_{\hbox{\st{eq}}} \sim 1.8\times  10^{-7}\,$Hz, which is due to the transition to the late MD era. $f_{\hbox{\st{eq}}}$ indicates the frequency corresponding to the emission around the matter-radiation equality time. We will elaborate the $f$-$T$ or $f$-$t$ correspondence later in Sec.~\ref{Sec:4-2}. Note that the $f^{-1/3}$ behavior was obtained by summing up to high oscillation modes $k \gg 10^5$ which was shown to be important for a MD background \cite{Blasi:2020wpy,Cui:2019kkd}. By only summing up to low $k$ modes ($k\ll 10^{5}$) it would be $f^{-1}$. The low end of the frequency spectrum has a cutoff corresponding to emission at the present time $t_0$, with a maximum point shortly before the ending of the spectrum at $f_0$.

By combining Eq.(\ref{Eq3-16}) and Eq.(\ref{Eq4-7}) we derive the following analytical approximation for global string SGWB spectrum in different $f$ regions, which shows the parametric dependence:
\begin{align}
\label{Eq5-1}
\Omega_{\hbox{\st{GW}}}(f)h^2 \simeq \left\{            
\begin{aligned}
\, & 5.1 \times 10^{-15} \left(\frac{\eta}{10^{15}\,\hbox{GeV}}\right)^{4} \left( \frac{f}{ f_{\eta}} \right)^{-1/3},\;\;\;\;\;\;\;\;\;\;\;\;\;\;\;\;\;\;\;\;\;\;\;\;\;\;\;\;\;\;\;\;\;\;\;\;\;\;\;\;\;\;\;\;\;\;\;\, \hbox{for} \; f> f_{\eta}  \\
 \, & 8.8\times 10^{-18} \left(\frac{\eta}{10^{15}\,\hbox{GeV}}\right)^{4} \hbox{ln}^3 \left[ \left( \frac{2}{\alpha f}\right)^{2} \frac{\eta}{t_{\rm eq}} \frac{1}{z_{\rm eq}^2  \sqrt{\xi}} \Delta_R^{1/2}(f) \right] \Delta_R(f), \;\;\;\; \hbox{for} \;  f_{\eta} > f> f_{\hbox{\st{eq}}} \\
\, & 2.9 \times 10^{-12} \left(\frac{\eta}{10^{15}\,\hbox{GeV}}\right)^{4} \left( \frac{f}{f_{\hbox{\st{eq}}}} \right)^{-1/3} , \;\;\;\;\;\;\;\;\;\;\;\;\;\;\;\;\;\;\;\;\;\;\;\;\;\;\;\;\;\;\;\;\;\;\;\;\;\;\;\;\;\;\;\;\;\;\, \hbox{for} \;f_0 < f< f_{\hbox{\st{eq}}}\\
\, & 0, \;\;\;\;\;\;\;\;\;\;\;\;\;\;\;\;\;\;\;\;\;\;\;\;\;\;\;\;\;\;\;\;\;\;\;\;\;\;\;\;\;\;\;\;\;\;\;\,\;\;\;\;\;\;\;\;\;\;\;\;\;\;\;\;\;\;\;\;\;\;\;\;\;\;\;\;\;\;\;\;\;\;\;\;\;\;\;\;\;\;\;\;\;\;\;\;\;\;\;\;\;\;\; \hbox{for} \; f< f_0
\end{aligned}
\right.
\end{align}
where $t_{\rm eq}$ and $z_{\rm eq}$ denote the time and redshift at the matter-radiation equality, respectively. $\Delta_R(f) $ accounts for the effect of varying the number of relativistic degrees of freedom, $g_*$ and $g_{*S}$, over time:
\begin{align}
\label{Eq: Delta_R}
\Delta_R(f) = \frac{g_*(f)}{g_*^0}\left( \frac{g_{*S}^0}{g_{*S}(f)} \right)^{4/3},
\end{align}
where $g_{*}(f)$ and $g_{*S}(f)$, are obtained by applying the $f$-$T$ relation which will be introduced later in Eq.(\ref{Eq4-1}), and the superscript $0$ indicates the values today. Note that here we focus on global strings associated with massless Goldstones, thus they are stable until the current time. In the case of axion strings with massive Goldstones, the string network would turn to domain walls and finally disintegrate around the transition time when Goldstones acquire masses. In that case, the GW spectrum would beget a cut with potentially distinct structure
around a characteristic $f_{\rm cut}$ that is larger than $f_0$. 

\subsection{GW frequency spectrum and experimental sensitivities}\label{sec:GWFSaES}
\begin{figure}[t]
\centering 
\includegraphics[width=0.75\textwidth]{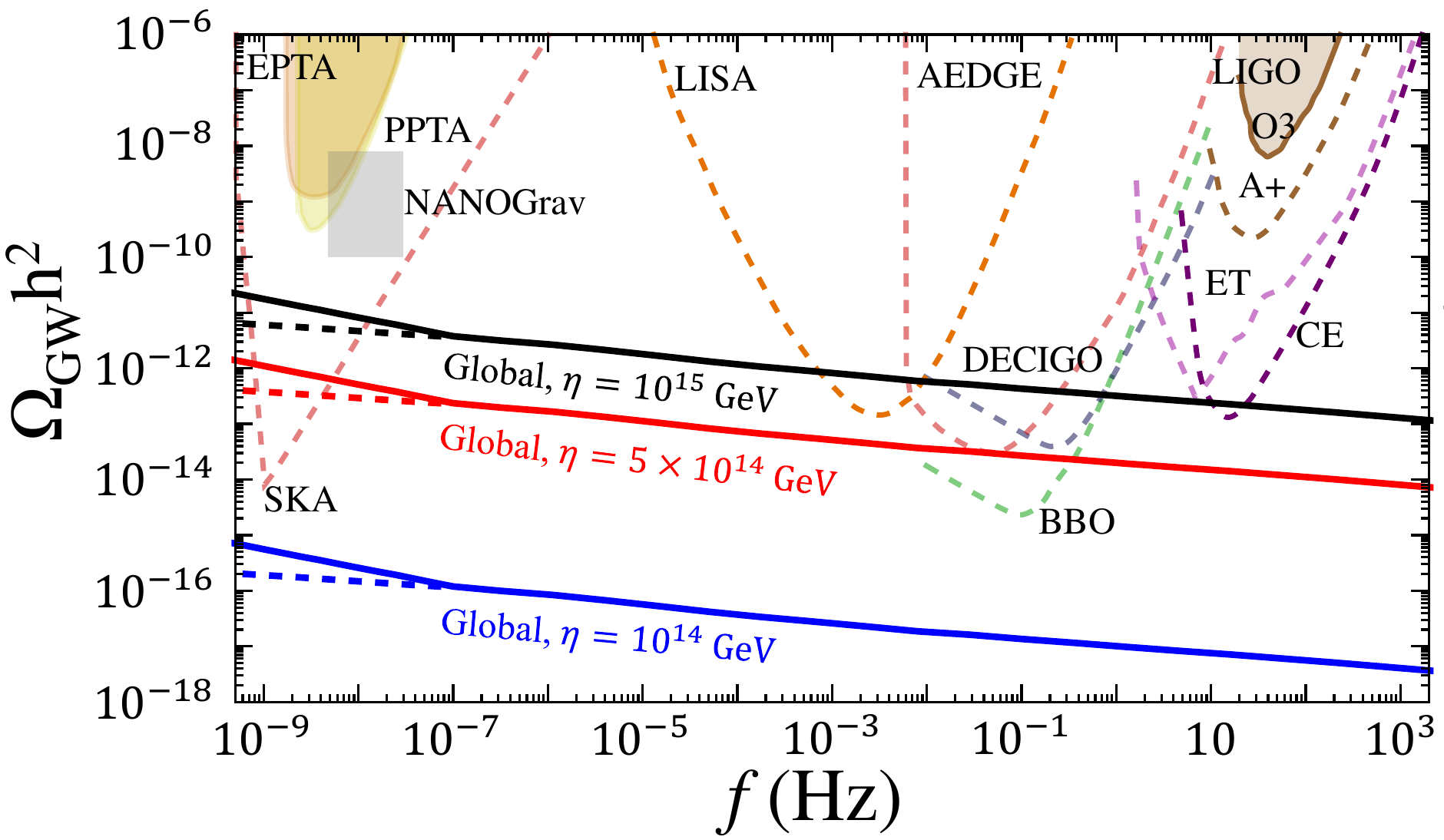} 
\caption{\label{F6} Gravitational wave spectrum from a global cosmic string network with $\alpha=0.1$, $F_{\alpha}=0.1$ for $\eta=10^{14}, \, 5\times 10^{14}, \, 10^{15}\,$GeV. The solid curves shown are the full results with standard cosmology, dashed lines show the contribution from emission during radiation domination. Exclusion limits or projected sensitivities with various GW experiments are also shown.}
\end{figure}

Fig.~\ref{F6} illustrates the SGWB signal originated from global strings based on our numerical results. We also include related experimental sensitivities: current constraints (solid lines) from LIGO \cite{TheLIGOScientific:2014jea,Thrane:2013oya,TheLIGOScientific:2016wyq,Abbott:2017mem} and European Pulsar Timing Array (EPTA) \cite{vanHaasteren:2011ni}, Parkes Pulsar Timing Array (PPTA) \cite{Lasky:2015lej,Shannon:2015ect, Blanco-Pillado:2017rnf}; the projected future sensitivities (dashed lines) with LIGO A+ \cite{Aasi:2013wya}, LISA \cite{Bartolo:2016ami}, DECIGO/BBO \cite{Yagi:2011wg}, AEDGE/AION \cite{Bertoldi:2019tck,Badurina:2019hst}, Einstein Telescope (ET) \cite{Punturo:2010zz,Hild:2010id}, Cosmic Explorer (CE) \cite{Evans:2016mbw}, and Square Kilometer Array (SKA) \cite{Janssen:2014dka}; as well as the region corresponding to the recent NANOGrav excess \cite{Arzoumanian:2020vkk,Arzoumanian:2021teu}. We can see that as expected from Eq.(\ref{Eq3-16}) the global string GW spectrum is sensitive to the symmetry breaking scale $\eta$. Experiments such as LISA, BBO and SKA can probe $\eta\gtrsim10^{14}$ GeV. Among the existing searches, PPTA gives the strongest constraint of $\eta \lesssim 2 \times 10^{15}\,$GeV. These sensitivities/constraints on $\eta$ may be improved/relieved with non-standard cosmology and alternative modelings, see discussions in Sec.~\ref{sec:DEUGS} and Sec.~\ref{sec:Diss}. Various intriguing interpretations of the recent NANOGrav excess as a SGWB signal have been considered \cite{Ellis:2020ena,Blasi:2020mfx,Neronov:2020qrl,DeLuca:2020agl,Buchmuller:2020lbh,Vagnozzi:2020gtf,Kohri:2020qqd,Ramberg:2020oct,Blanco-Pillado:2021ygr}. In particular, \cite{Gorghetto:2021fsn} and \cite{Ramberg:2020oct} investigated the possibility of fitting the NANOGrav signal with GWs from QCD axion strings or general ALP strings. The former \cite{Gorghetto:2021fsn} found that the GW amplitude hinted by the NANOGrav data requires $f_a \gtrsim 10^{15}\,$GeV which is in conflict with bound on $\Delta N_{\hbox{\st{eff}}}$ from BBN and CMB data, given that the axions are emitted as radiation from the strings. Nevertheless, the latter suggests that a non-standard cosmological history may improve the fit \cite{Ramberg:2020oct}. In our independent check by including high $k$ modes in the summation, we find that the GW frequency spectrum follows a power-law $f^{-1/3}$ in the range of $f\leq f_{\rm eq}$ (defined before Eq.(\ref{Eq5-1})), and with $4.3\times10^{15}\,$GeV$\,\leq \eta \leq 6.1\times10^{15}\,$GeV, global strings can lead to a good 1-$\sigma$ fit to the NANOGrav 12.5-year data \cite{Arzoumanian:2020vkk}. However, as also discussed in \cite{Arzoumanian:2020vkk,Ellis:2020ena}, such a spectrum with a gentle slope is in tension with previous bounds from PPTA \cite{Shannon:2015ect, Lasky:2015lej}, EPTA \cite{vanHaasteren:2011ni}, and NANOGrav 11-year data \cite{Arzoumanian:2018saf}. Such a tension may be eased by re-analyzing the data sets using different choices of the red noise model \cite{Hazboun:2020kzd} which is being investigated. Variations to the standard theoretical assumptions may allow a viable interpretation of the NANOGrav signal as originated from a global/axion string network, consistent with PPTA data and $\Delta N_{\hbox{\st{eff}}}$ constraints, which we will explore in future study. In Sec.~\ref{sec:RDGG}, we will discuss the $\Delta N_{\hbox{\st{eff}}}$ bound on Goldstone and GW emissions in detail. Other relevant constraints on the global $U(1)$ breaking scale $\eta$ include inflation scale and CMB anisotropy bound, which were discussed in \cite{Chang:2019mza,Lopez-Eiguren:2017}, also pointing to $\eta\lesssim O(10^{15})$ GeV. CMB polarization data potentially yields stronger bound on GW in the frequency range of $10^{-17}-10^{-14}$ Hz \cite{Lasky:2015lej,Smith:2005,Namikawa:2019tax}. Nevertheless, in Sec.~\ref{Sec:4-2} we will demonstrate that this latter constraint does not apply to our case following the introduction of the $f$-$T$ relation. 

\textit{Comparison with literature:}\\
GW signals from a global string network have also been recently investigated by simulation approaches based on a Nambu-Goto effective theory \cite{Gorghetto:2021fsn} or field theory for global defects \cite{Figueroa:2020lvo}. Our results agree with others' on some general features such as $\Omega_{\hbox{\st{GW}}}\propto \eta^4$, but differ in details. \cite{Gorghetto:2021fsn} simulated the global string network in a radiation background during a very early stage of evolution, i.e. $N\lesssim 7$-$8$, and extrapolated the linear growth of $\xi \propto N$ to high $N$ when computing the GW spectrum. They agree with our finding that the global strings can lead to detectable GW signals, but found that the GW spectrum scales as $\Omega_{\hbox{\st{GW}}} \propto \eta^4 N^4$, instead of $\eta^4 N^3$ as found in our analysis (see Eq.(\ref{Eq3-16}) in Sec.~\ref{sec:RDGG}). The $N^3$ dependence we found results from the prediction of the conventional scaling VOS model. 
The difference may be resolved if the loop emission factor $C_{\hbox{\st{eff}}}$ in the VOS model is not (nearly) a constant but evolves as $C_{\hbox{\st{eff}}}\propto N$ (see Eq.(\ref{Eq: Ceff}). We further discuss the effect of such a non-scaling behavior or deviation from the conventional VOS in Sec.~\ref{Sec:Non-scaling solution}. On the other hand, \cite{Figueroa:2020lvo} found that the GW spectrum asymptotes to an exact scale invariant form, and the amplitude of the signal is below the prediction by both our method and by \cite{Gorghetto:2021fsn}. The possible explanations for this discrepancy was suggested in \cite{Chang:2019mza, Gorghetto:2021fsn}, while further investigation is certainly needed to fully resolve this issue.

\subsection{Comparison with GWs from NG strings, relic densities of GWs and (massless) Goldstones}\label{sec:RDGG}

\begin{figure}[t]
\centering
\includegraphics[width=0.8\textwidth]{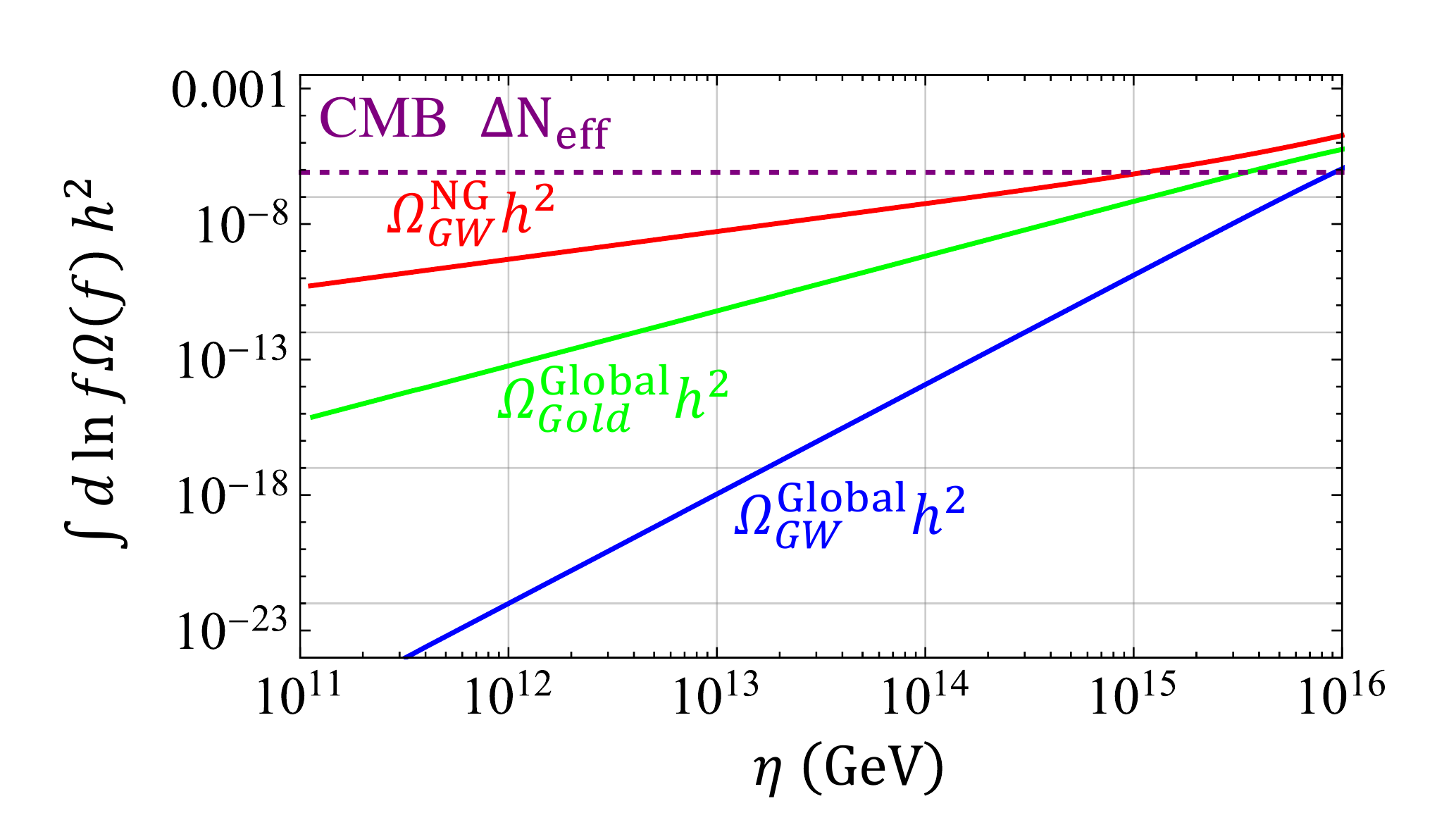} 
\caption{\label{F5} The total relic densities (integrated over $f$) of GWs from a NG string network (red), of GWs from a global string network (blue), and of massless radiation-like Goldstone bosons from a global string network (green), as functions of the symmetry breaking scale $\eta$ (related to the string tension $\mu$). The purple dashed line shows the constraint on extra radiation energy density by CMB data: $\Delta \hbox{N}_{\rm eff} \lesssim 0.2$ \cite{Aghanim:2018eyx} or $\int d(\hbox{ln}f)\Delta\Omega_{\hbox{\st{rad}}}h^2 \lesssim  8.1 \times 10^{-7}$ \cite{Aghanim:2018eyx, Henrot-Versille:2014jua}, which requires $\eta \lesssim 3.5\times 10^{15}\,$GeV. }
\end{figure}

In this subsection, we give a simple estimate for the relic density of GWs from global strings which captures key parametric dependence, and compare it with that for NG strings. This can help us gain insights about the detectability of the GW signal from global strings. 
 In addition, while in this work we focus on GW radiation from global strings, it is important to better understand Goldstone emission which is the dominant radiation mode in this case. We thus also present a parametric estimate for the relic density of the emitted Goldstones and compare it with GW emission. As shown in Fig.~\ref{F5}, with these analyses we can find the constraint on $\eta$ considering the upper limit on extra radiation energy density $\Delta N_{\rm eff}$ from BBN/CMB data. The emitted GWs can potentially affect CMB observables in other ways (e.g. CMB polarization) and lead to other constraints, which we will discuss later in Sec.~\ref{Sec:4-2}. As mentioned earlier, in this work we focus on the simple case with massless Goldstones and our discussion about Goldstone emission is illustrative and concise. Nevertheless, some key insights can be applied to axion strings where the Goldstones are massive as potential dark matter candidates. We leave a detailed discussion on Goldstone radiation and its impact on axion DM physics for future work.

A key difference between the dynamics of a global and a NG string network is that the global string loops are rather short-lived due to the strong Goldstone emission rate. We consider a loop formed at time $t_i$ which decays away at time $t_r\equiv \gamma_r(t_i)t_i$, where we have adopted a unified notation for the cases of NG and global string loops for an easy comparison: $r = \{\hbox{NG}, \, \hbox{global}\}$. Using Eq.(\ref{Eq3-2}) we find the following expression for estimating the lifetime parameter $\gamma(t_i)$ for the two cases:
\begin{align}
\label{Eq3-9}
\gamma_r (t) \equiv \frac{\alpha + \Gamma G \mu + \kappa}{\Gamma G \mu +\kappa} &\simeq \left\{     
\begin{aligned}
&\frac{\alpha}{\Gamma G \mu} \simeq 5\times 10^{10} \left( \frac{\eta}{10^{12}\hbox{GeV}} \right)^{-2}  \;\;\;\;\;\;\;\, \hbox{NG String ($\kappa=0$)} \\
&\frac{\alpha+\kappa}{\kappa} \simeq 2 \;\;\;\;\;\; \;\;\;\;\;\; \;\;\;\;\;\; \;\;\;\;\;\; \;\;\;\;\;\;\;\; \;\;\;\;\;\; \;\;\; \hbox{Global String}
\end{aligned}
\right.
\end{align}
where
\begin{align}
\label{Eq: kappa}
\kappa\equiv \frac{\Gamma_a}{2\pi N}.
\end{align}
Our ansatz of $\alpha\sim0.1\gg\Gamma G\mu$ is applied to derive the final results. The lifetime of a loop formed at time $t_i$ with an initial length of $\alpha t_i$ can then be estimated as $\tau_r = (\gamma_r(t_i)-1)t_i$. Recent simulations support our estimates of global string loop's lifetime \cite{Saurabh:2020pqe,Gorghetto:2020qws}. Due to the time dependence of global string tension (i.e. the $N$-dependence), $\kappa$ varies in the range of $ 0.6 \lesssim 1/\kappa \lesssim 10 $ throughout the expansion history of universe.  Therefore, the global strings are short-lived and are expected to decay in about one Hubble time after its formation (but the lifetime is still sufficient to yield detectable GWs with large $\eta$). In contrast, as can be seen from Eq.(\ref{Eq3-9}), the NG string loops generally survive a much longer time after formation. Due to this drastic difference in loop lifetime, with the same parameters such as $\eta$ and loop distribution function, GWs from a global string network on average experience a larger redshift effect after emission, which contributes to a suppressed GW amplitude (along with the suppression effect due to the Goldstone dominance) and shifts the spectrum towards lower frequencies. These considerations help us understand the numerical results as shown in Fig.~\ref{F3}.

We now estimate the relic densities of GW and Goldstone emitted from global strings. As mentioned in Sec.~\ref{sec:SGSN} the formulation for GW calculation given in Eq.(\ref{Eq3-6}) and Eq.(\ref{Eq3-7}) can be applied to the Goldstone case with the replacements of $\Gamma \to \Gamma_a,$ and $\Gamma G \mu^2 \to \Gamma_a \eta^2$. (based on Eq.(\ref{Eq3-1})). We can then express the total relic densities (integrated over $f$) of GWs and Goldstones from global string radiation in the following unified form:
\begin{align}
\label{Eq3-11}
\Omega_{\beta} = \int d\left( \hbox{ln}f \right) \Omega_\beta (f), \;\;\;\;\;\hbox{with} \;\; \beta = \{\hbox{GW}, \, \hbox{Gold}\}.
\end{align}
Our numerical results of the relic energy densities are illustrated in Fig.~\ref{F5} as functions of symmetry breaking scale $\eta$, along with $\Omega_{\rm GW}$ from NG strings for comparison. The upper limit on the total relic radiation energy density from the CMB data is also shown \cite{Smith:2006nka,Henrot-Versille:2014jua,Aghanim:2018eyx}. One can see that the constraint is dominantly driven by the emission of radiation-like Goldstones, which requires $\eta\lesssim 3.5\times10^{15}$ GeV, while for GWs alone the constraint is relaxed to $\eta\lesssim  9\times 10^{15}$ GeV. In comparison, with a non-scaling solution as suggested in Eq.(\ref{Eq2-6-2}) this CMB constraint on $\eta$ would be tighter:  $\eta \lesssim 9 \times 10^{14}\,$GeV \cite{Gorghetto:2021fsn}, as the total energy of the string network would increase relative to the scaling scenario (see Sec.~\ref{Sec:Non-scaling solution} for more related discussion).

Next we further discuss the parametric dependence of $\Omega_{\rm GW}$ and $\Omega_{\rm Gold}$ for global strings and compare with $\Omega_{\rm GW}$ for NG string. For a fair comparison, we assume that the symmetry breaking scale $\eta$ and the string network evolution  parameters such as the long string number density $\xi$ and loop size $\alpha$ are the same for the NG and global string network under consideration. Then we consider $\Omega^{\rm Global}_{\rm GW}$, $\Omega^{\rm Global}_{\rm Gold}$ and $\Omega^{\rm NG}_{\rm GW}$ as observed at a time parametrized by $N\equiv \hbox{ln}(L\eta)$. Based on simple analytic estimates checked with numerical fitting, we find the following relations: 

\begin{align}
\label{Eq3-14}
\Omega^{\hbox{\scriptsize{NG}}}_{\hbox{\st{GW}}} : \Omega_{\hbox{\scriptsize{Gold}}}^{\rm Global} : \Omega^{\hbox{\scriptsize{Global}}}_{\hbox{\st{GW}}} \, \simeq & \; 1 :  N\sqrt{\frac{\Gamma G\mu}{\alpha}} : N\sqrt{\frac{\Gamma G\mu}{\alpha}}  \frac{\Gamma G \mu}{\Gamma_a/(2\pi N)},
\end{align}
where $\frac{\alpha}{\Gamma G\mu}$ is the lifetime parameter for NG string, $\gamma_r^{\rm NG}$, as defined in Eq.(\ref{Eq3-9}), which accounts for the aforementioned difference in redshift effects between global and NG case, and the square-root of $\frac{\Gamma G\mu}{\alpha}$ is due to the redshift of the GW energy $\propto a(t) \propto t^{1/2}$; the $N$ factors account for the $\log$ enhanced string tension for global strings; $\frac{\Gamma G \mu}{\Gamma_a/(2\pi N)}$ represents the different energy loss rates to GWs vs. to Goldstones. We also find the following key parametric dependencies (focusing on $\eta$ and $N$) for each of these $\Omega$'s:
\begin{align}
\label{Eq3-16}
\Omega^{\hbox{\st{NG}}}_{\hbox{\st{GW}}} \propto \eta, \;\;\;\;\;\;
\Omega_{\hbox{\st{Gold}}}^{\rm Global} \propto \eta^2 N, \;\;\;\;\;\;
\Omega^{\hbox{\st{Global}}}_{\hbox{\st{GW}}} \propto \eta^4  N^3,
\end{align}

The $\eta$ dependence of GWs from NG strings that we found agrees with earlier literature \cite{Vilenkin:2000jqa,Cui:2018,Cui:2017,Vilenkin:1981bx,Vachaspati:1984gt,Olum:1999sg,Figueroa:2012kw,Blanco-Pillado:2017rnf,Blanco-Pillado:2017oxo,Ringeval:2017eww}, and $\Omega_{\hbox{\st{GW}}}^{\hbox{\st{Global}}}\propto \eta^4$ agrees with two most recent independent simulations \cite{Gorghetto:2021fsn} and \cite{Figueroa:2020lvo}. Nevertheless, the $N$-dependence of the scaling solution of long string number density $\xi$ in the VOS model (see Eq.(\ref{Eq2-5})) disagrees with some of the simulation results which suggest a logarithmic increase in $\xi$ based on low $N$ data \cite{Gorghetto:2021fsn}. The effect of a non-scaling $\xi$ persisting till late times, e.g. $N >70$, will be discussed in  Sec.~\ref{Sec:Non-scaling solution}, including a comparison with the result in \cite{Gorghetto:2021fsn}.

\section{Probing the Early Universe with SGWB Spectrum from Global Strings}
\label{sec:DEUGS}
In this section we investigate how the SGWB spectrum from a global string network would alter if the cosmic history and particle content of the early Universe differ from the standard scenario which we assumed in Sec.~\ref{sec:GWSGCS}. This in turn allows us to use such GW signals to test the standard paradigms and probe the dynamics of the early Universe well before BBN. Such an idea of using GWs for \textit{cosmic archaeology} was proposed and developed in the context of NG strings \cite{Cui:2017,Cui:2018}. The situation with global strings bear similarities with that of NG strings, yet with significant differences. In the following, we will demonstrate our findings and make comparison with NG string results.

\subsection{The connection between the observed GW frequencies and emission times}\label{Sec:4-2}
In the context of NG strings, the frequency-temperature ($f$-$T$) correspondence during a RD era was derived in \cite{Cui:2018}, and serves as the foundation of cosmic archaeology with the $f$ spectrum of GWs from strings. The analogous relation for global strings can be derived following the same method. Nevertheless, the derivation can be greatly simplified in this case. As explained in Sec.~\ref{sec:RDGG} (Eq.(\ref{Eq3-16})), a key difference between NG and global string loop dynamics is that, global string loops decay away within $\sim1$ Hubble time after formation due to the strong Goldstone emission rate. Therefore, the timescale when the GW emission from a loop occurs is approximately the same as the loop's formation time, i.e. $\tilde{t} \sim t_i$ (Eq.~(\ref{Eq3-2})). For an estimate, it suffices to focus on the $k =1$ mode which we find to be the dominant one in the cases of interest. With this understanding and following the calculation in Sec.~\ref{sec:SGSN}, we find that a specific $f_{\Delta}$ band observed today relates to a particular emission temperature $T_{\Delta}$ in the following way: 
\begin{align}
\label{Eq4-1}
\notag f_\Delta & \simeq \frac{2}{\ell (\tilde{t})}\frac{a(t_{\Delta})}{a(t_0)} = \frac{2}{\alpha z_{\hbox{\st{eq}}} t_{\hbox{\st{eq}}} T_{\hbox{\st{eq}}}} \left[ \frac{g_*(T_\Delta)}{g_*(T_{\hbox{\st{eq}}})} \right]^{1/4} T_\Delta\\
&\simeq (3.02 \times 10^{-6} \, \hbox{Hz}) \left( \frac{T_\Delta}{1\,\hbox{GeV}} \right) \left( \frac{\alpha}{0.1} \right)^{-1} \left[ \frac{g_*(T_\Delta)}{g_*(T_{\hbox{\st{eq}}})} \right]^{1/4},
\end{align} 
where the loop size at the emission time $\ell (\tilde{t})  \simeq \alpha t_{i} \equiv \alpha t_{\Delta}$ (see Eq.(\ref{Eq3-2})), $z_{\hbox{\st{eq}}} \simeq 3387$ is the redshift at the matter-radiation equality, and $t_{{\hbox{\st{eq}}}}$, $T_{\hbox{\st{eq}}}$ are the corresponding time and temperature, respectively. Note that $f_\Delta$ linearly depends on $T_\Delta$, but is insensitive to the symmetry breaking scale $\eta$, unlike in the case of local strings. Eq.~(\ref{Eq4-1}) applies to RD era, while $f$-$T$ relation varies with background cosmology, which we will discuss in Sec.~\ref{sec:PNPOCE}. A departure from the standard cosmology at $T_\Delta$ would thus imprint itself in the GW spectrum around the corresponding $f_\Delta$.\\

\begin{figure}[t!]
\centering 
\includegraphics[width=0.79\textwidth]{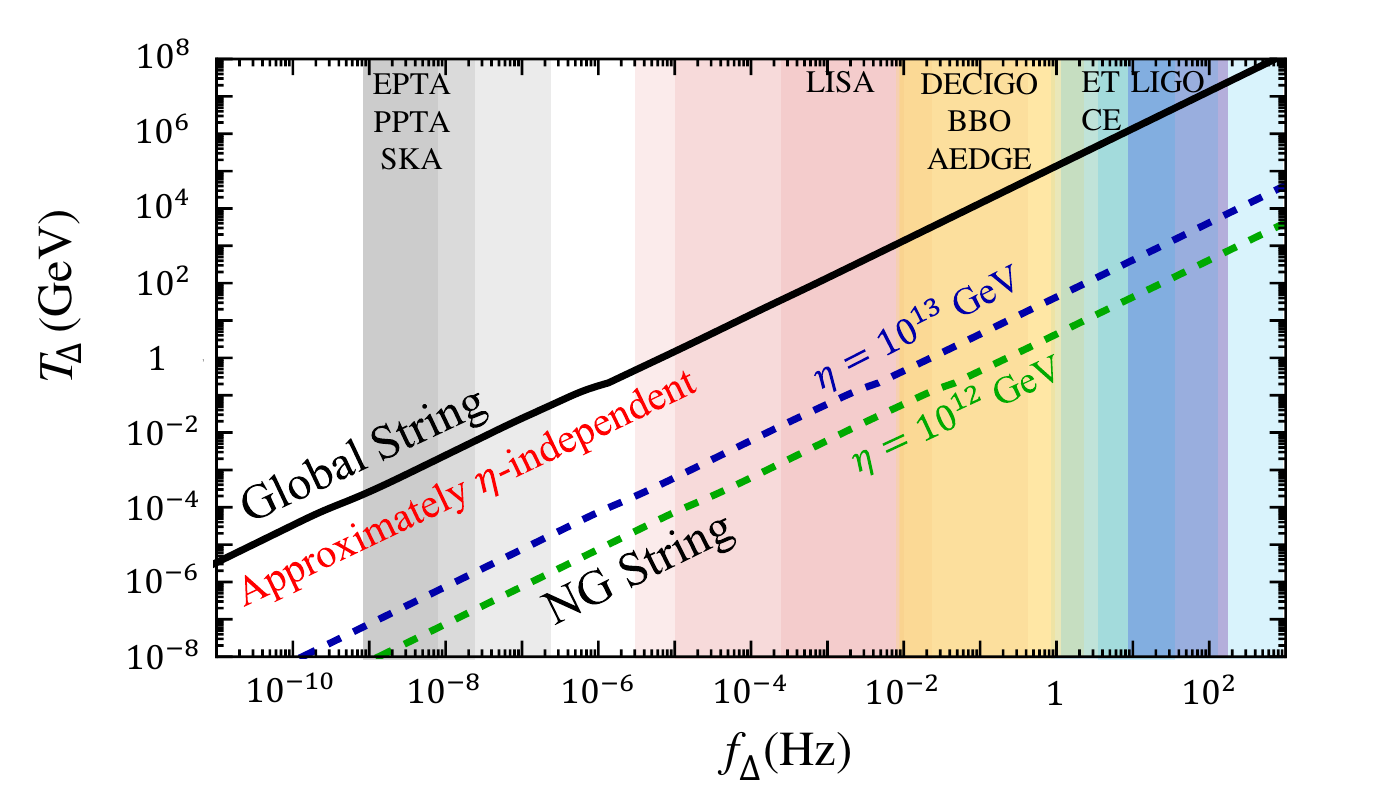} 
\caption{\label{F8} Frequency $f_\Delta$ where the GW spectrum from cosmic strings would be altered due to a transition to a non-standard cosmology at $T_\Delta$ (Eq.(\ref{Eq4-1})): the comparison between the results for global strings (the upper-left black line) and NG strings \cite{Cui:2018,Cui:2017} (the lower-right dashed lines). The relevant experimental sensitivities are also shown in different colors, where the darkest bands indicate peak sensitivities. This illustrated the $f_\Delta-T_\Delta$ relation given in the main text.}
\end{figure}

\begin{figure}[t!]
\centering
\includegraphics[width=0.7\textwidth]{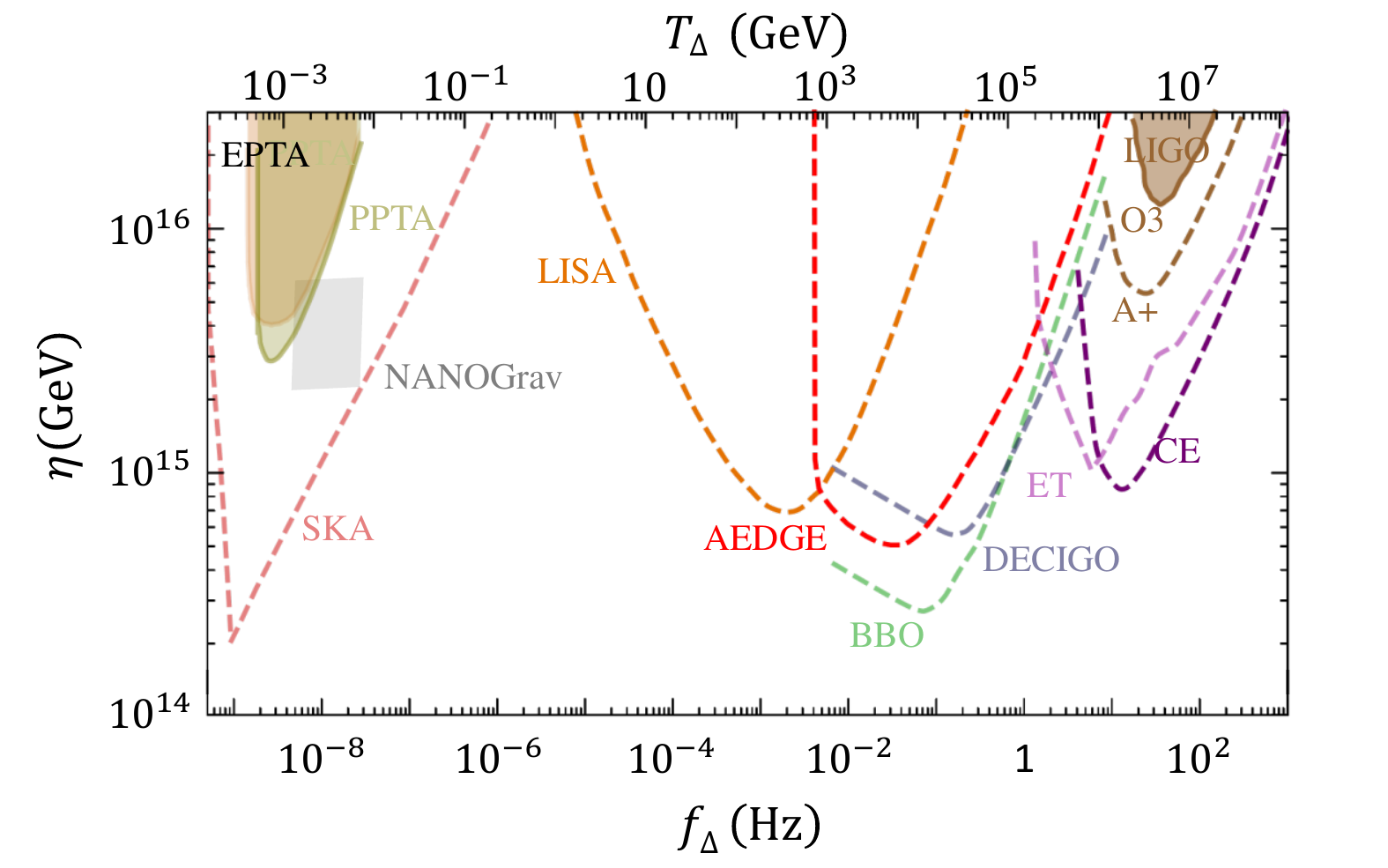} 
\caption{\label{F7} Another illustration for $f_\Delta$-$T_\Delta$ relation for GW frequency spectrum from global strings, where the experimental sensitivities to $\eta$ are shown.}
\end{figure}

In Fig.~\ref{F8} we illustrate the $f$-$T$ relation derived for SGWB spectrum from global strings, in comparison with the recent results for NG strings \cite{Cui:2018,Cui:2017}. There are two major differences between the two cases: $f$-$T$ correspondence for NG strings has $\eta$-dependence while for global strings it is almost independent of $\eta$ which makes it more robust in a way; for the same $f$ the corresponding emission $T$ is much earlier for global strings than for NG strings. Both these differences originate from the aforementioned fact that global string loops decay shortly after formation and their resultant GW signal observed in a certain $f$ band has undergone longer period of redshift after emission (relative to its NG counterpart). Note that due to the current bounds from LIGO and PPTA, $\eta$ for NG strings is constrained as $\eta \lesssim 1.89 \times 10^{13}\,$GeV \cite{Cui:2018, Blanco-Pillado:2017rnf}. If the recent NANOGrav excess indeed originates from NG cosmic strings, it favors $\eta\simeq 3-5 \times 10^{13}\,$GeV \cite{Ellis:2020ena,Blasi:2020mfx}. According to Fig.~\ref{F8} these constraints/potential signal implies that GW spectrum from NG strings can reach up to $T\sim10^4$ GeV (with ET and CE). In contrast, as shown in Fig.~\ref{F8} global strings can probe much earlier cosmic history, up to $T\sim10^8$ GeV. As discussed in Sec.~\ref{sec:GWFSaES}, $\eta \gtrsim 10^{14}\,$GeV is needed to be within experimental sensitivity reach in terms of $\Omega_{\rm GW}$, while other constraints require $\eta \lesssim O(10^{15})\,$GeV. Fig.~\ref{F7} illustrates the $f$-$T$ relation for global strings in a different manner where the sensitivity to $\eta$ is explicitly shown. As demonstrated in Fig.~\ref{F7}, global string GWs can trace the cosmic history over a rather wide range in time: up to $T \sim 10^8$ GeV (with  ET and CE) and down to $T \sim 10^{-4}\,$GeV (with PPTA and SKA) which intriguingly corresponds to the beginning of the BBN era. 


\begin{figure}[t]
\centering
\includegraphics[width=0.9\textwidth]{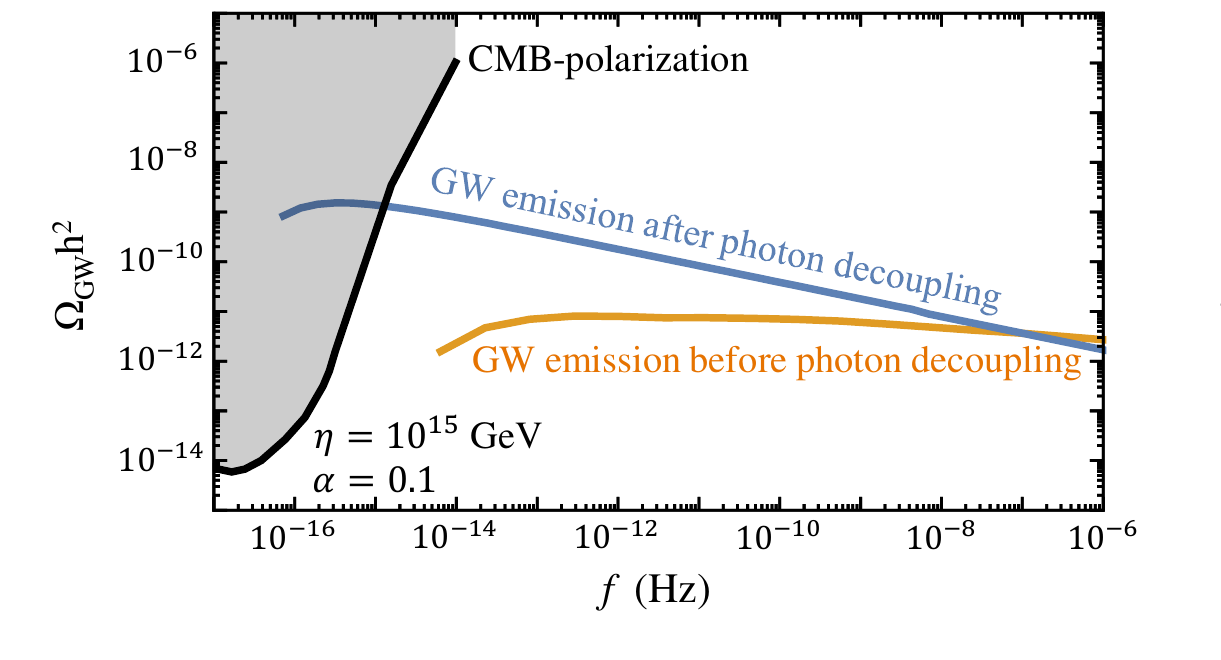} 
\caption{\label{Fig:CMB} GW spectrum from a global string network decomposed into contributions from before and after photon decoupling, which demonstrates how CMB polarization constraint is safely evaded (see main text for details). In the example shown, $\eta=10^{15}\,$GeV ($G\mu = 10^{-11}$) .}
\end{figure}

The $f$-$T$ relation as we have elaborated can also help us understand why the global string scenario safely evades the potentially strong bound on $\Omega_{\rm GW}$ in the range of $f\sim10^{-17}-10^{-14}$ Hz by the CMB polarization data \cite{Smith:2006nka,Lasky:2015lej,Smith:2005,Namikawa:2019tax}. 
The $f$-$T$ relation in Eq.(\ref{Eq4-1}), together with the observation that global string loops decay in one Hubble time, indicate that the SGWB signal below a certain $f$ range could not be generated until after a certain time or below a certain $T$. In Fig.~\ref{Fig:CMB} we illustrate the constraints from CMB polarization data, and the decomposed contributions to a SGWB induced by global strings: the signal in the low $f$ range of $f\sim10^{-17}-10^{-14}$ Hz in fact is not populated until after the photon decoupling, thus is not present at the CMB epoch to be subject to the constraint. One can also simply estimate $f$ corresponding to the photon decoupling $T_{\gamma} \sim 0.3\,$eV using Eq.(\ref{Eq4-1}), and confirm that GWs with $f\lesssim10^{-15}$ Hz is emitted afterwards.

\subsection{Probing new phases of cosmological evolution}\label{sec:PNPOCE}
According to the standard thermal history, the Universe is radiation dominated starting from the end of inflation all the way down to the matter-radiation equality at $z_{\rm eq}\sim3000$. Nevertheless, so far there is no data evidence to support this assumption for the epoch prior to the BBN time, i.e. the primordial dark age. On the other hand, recently there has been substantial interest to consider well-motivated non-standard cosmology scenarios, where the standard RD era transits to a different phase at some point in the early Universe, such as EMD or kination. An EMD era can be due to the temporary domination of a long-lived massive particle or oscillations of a scalar moduli field \cite{Huey:2000jx}. More generic possibilities arise in models where a scalar field $\phi$ oscillates in a polynomial potential $V(\phi) \propto \phi^N$, characterized by an averaged equation of state $w = (N-2)/(N+2)$.  In the limit $N \to \infty$, we have $n = 6$ in Eq.(\ref{Eq2-4}) which is called kination phase, as the kinetic energy of the scalar dominates. Kination can generally arise in inflation \cite{Salati:2002md}, quintessence, dark energy \cite{Chung:2007vz}, and axion-like particle (ALP) models with varying power of sin-Gordon potential \cite{Poulin:2018dzj} or with a non-zero initial field velocity  \cite{Chang:2019tvx,Co:2019jts}. In order to retain the successful predictions of BBN theory, for all these scenarios the Universe needs to settle to RD 
before the BBN time $T_\Delta \sim 5\,$MeV. 

It is thus intriguing to see how the SGWB from global strings would alter in a non-standard cosmology and the related implication for detections. From another perspective, similar to the finding in the context of NG strings, SGWB from global strings thus opens up the possibility of probing the early Universe during the primordial dark age that may not be directly accessible otherwise. This allows us to test the standard assumption about cosmology while uncovering potential deviations. The base of this method lies in the $f$-$T$ relation during RD (Eq.(\ref{Eq4-1})) which allows us to relate a deviation from the standard prediction for the SGWB frequency spectrum to a time point in history where RD transits to a new (earlier) phase. To calculate $\Omega_{\rm GW}(f)$ with a non-standard cosmology background, we follow the method given in \cite{Chang:2019mza,Cui:2018}: we assume that the Universe transits from RD to a new equation state parametrized by $n$ (Eq.(\ref{Eq2-4}): $\rho\propto a^{-n})$ at $T_\Delta$, and match the energy density at $T_\Delta$ for a smooth transition. $\Omega_{\rm GW}(f)$ can then be calculated using Eq.(\ref{Eq3-6}) with the input of a non-standard evolution of $a(t)$. 

We therefore expect that a non-standard cosmology leads to a modified GW spectrum in the high frequency region starting round $f_\Delta$ corresponding to the transition in cosmic history occurring around $T_\Delta$ (see Eq.(\ref{Eq4-1})). Numerically, we found that $\Omega_{\hbox{\st{GW}}}(f)$ can be parametrized in the following way in large $f$ region for general cosmologies (parametrized by $n$):
\begin{align}
\label{Eq4-2}
\Omega_{\hbox{\st{GW}}} (f ) \propto \left\{            
\begin{aligned}
&f^{\frac{8-2n}{2-n}} \, \hbox{log}^3 \Bigg\{ \frac{\eta}{\sqrt{\xi}} \left[  \left( \frac{20}{ f}\right)^{2} \frac{1}{t_{eq}} \frac{t_\Delta^{1-4/n}}{z_{eq}^2 } \Delta_R^{1/2}(f) \right]^{1/(2-4/n)} \Bigg\}, \;\;\;\;\;\;\;\;\;\;\;\;\;\;\;\;\;\hbox{for}\,\,\, n \geq \frac{26}{7} , \\
&f^{-\frac{1}{3}},\;\;\;\;\;\;\;\;\;\;\;\;\;\;\;\;\;\;\;\;\;\;\;\;\;\;\;\;\;\;\;\;\;\;\;\;\;\;\;\;\;\;\;\;\;\;\;\;\;\;\;\;\;\;\;\;\;\;\;\;\;\;\;\;\;\;\;\;\;\;\;\;\;\;\;\;\;\;\;\;\;\;\;\;\;\;\;\;\;\;\;\;\;\;\;\;\, \hbox{for} \,\,\, n < \frac{26}{7}, 
\end{aligned}
\right.
\end{align}
where $t_\Delta$ is the time corresponding to the temperature $T_\Delta$, and we have assumed $\alpha=0.1$. Eq.~(\ref{Eq4-2}) shows that $\Omega_{\hbox{\st{GW}}} (f \gg f_\Delta) \propto f^{+1} $ for  kination ($n = 6$) and $\propto f^{-1/3}$ for MD ($n = 3$). Note that the validity of the VOS model approach requires $n>2$, otherwise both $\xi$ and $\bar{v}_\infty$ would me imaginary valued according to Eq.~\ref{Eq2-5}. Another caveat is that, at sufficiently large $f\gtrsim f_\eta$ such that $\log(...)\sim 1$ or $N\sim O(1)$ (corresponding to the very early stage after the string network formation), $\Omega_{\rm GW}$ would universally fall as $\Omega_{\rm GW} \propto f^{-1/3}$, for different background cosmologies. 

\begin{figure}[t]
\centering 
\includegraphics[width=1\textwidth]{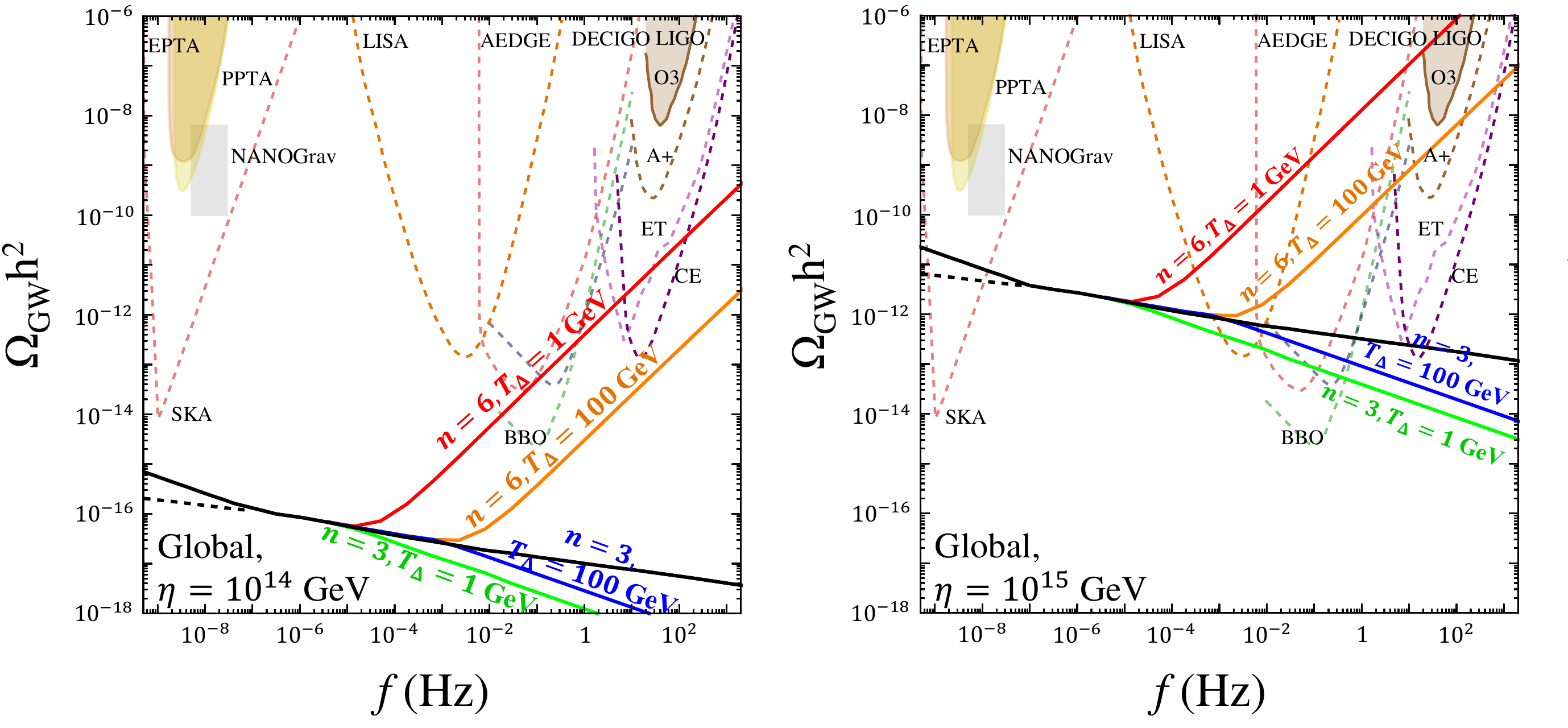} 
\caption{\label{F9} Gravitational wave spectrum from a global cosmic string network with $\alpha=0.1$, $F_{\alpha}=0.1$ for $\eta=10^{14}$ (left) and $10^{15}\,$GeV (right). The solid black lines show the GW spectrum with the standard cosmological evolution. The colored lines show the results with an EMD ($n=3$) or kination ($n=6$) that ends and restores the late RD era at the temperature $T_\Delta = 10\,$GeV or $10^2\,$GeV. The sensitivities of related GW experiments are also shown.}
\end{figure}

In Fig.~\ref{F9} we show our numerical results for benchmark examples of GW spectrum from a scaling global cosmic string network with a non-standard cosmology background such as kination or EMD, contrasted by the standard prediction shown in solid black line. We can see that compared to standard cosmology, with the presence of an EMD phase $\Omega_{\rm GW}(f)$ falls faster towards higher $f$, making it harder to observe in that $f$ range. On the other hand, the spectrum rises above the standard prediction at high $f$ in case of an early kination phase, leading to a stronger signal. The kination case thus is more subject to existing constraint from LIGO. In general, the LIGO O3 constraint on $T_\Delta$ can be expressed as
\begin{align}
\label{Eq4-3}
\left(\frac{T_\Delta}{100\,\hbox{GeV}} \right)^{-1} \left( \frac{\eta}{10^{15}\,\hbox{GeV}} \right)^{4} \lesssim 1,
\end{align}
where we have dropped the logarithmic dependence from Eq.(\ref{Eq4-2}) for a simple estimate. The LIGO constraint can be relaxed if the duration of kination is short enough so that it transits to other phases (e.g. RD, EMD or vacuum energy domination) at a time corresponding to an $f$ band below LIGO reach. In fact a sufficiently short span of kination epoch is also required to satisfy CMB/BBN bound on extra radiation density as we reviewed earlier in Sec.~\ref{sec:RDGG}. With these motivations, in the following we further consider a concrete example with two stages of transitions where kination is preceded by an earlier RD era and investigate the constraints on the model due to the CMB/BBN bound on $\Delta N_{\rm eff}$ (both Goldstone and GW). 

\noindent\textit{A two-stage transition scenario with kination:}\\
We assume that at $T=T_{\Delta 2}>T_{\Delta 1}$ kination transits to an early RD era (note: not the later standard RD era). Such a scenario can be realized if, for instance, a dominating radiation species decays to kination particles around $T_{\Delta 2}$. Other possibilities of exiting kination at high $T$ exist, e.g. by a vacuum energy dominated phase such as inflation. However, a long period of vacuum energy domination would dilute the overall GW signal significantly \cite{Guedes:2018afo, Cui:2019kkd}. An alternative is to have a short duration of vacuum energy domination (mini-inflation) which then transits to an early RD. Kination can also be preceded by a dominating matter-like species that decay to kination particles. Here we choose to consider the simple scenario of RD-kination-RD for illustration.

\begin{figure}[t]
\centering 
\includegraphics[width=0.48\textwidth]{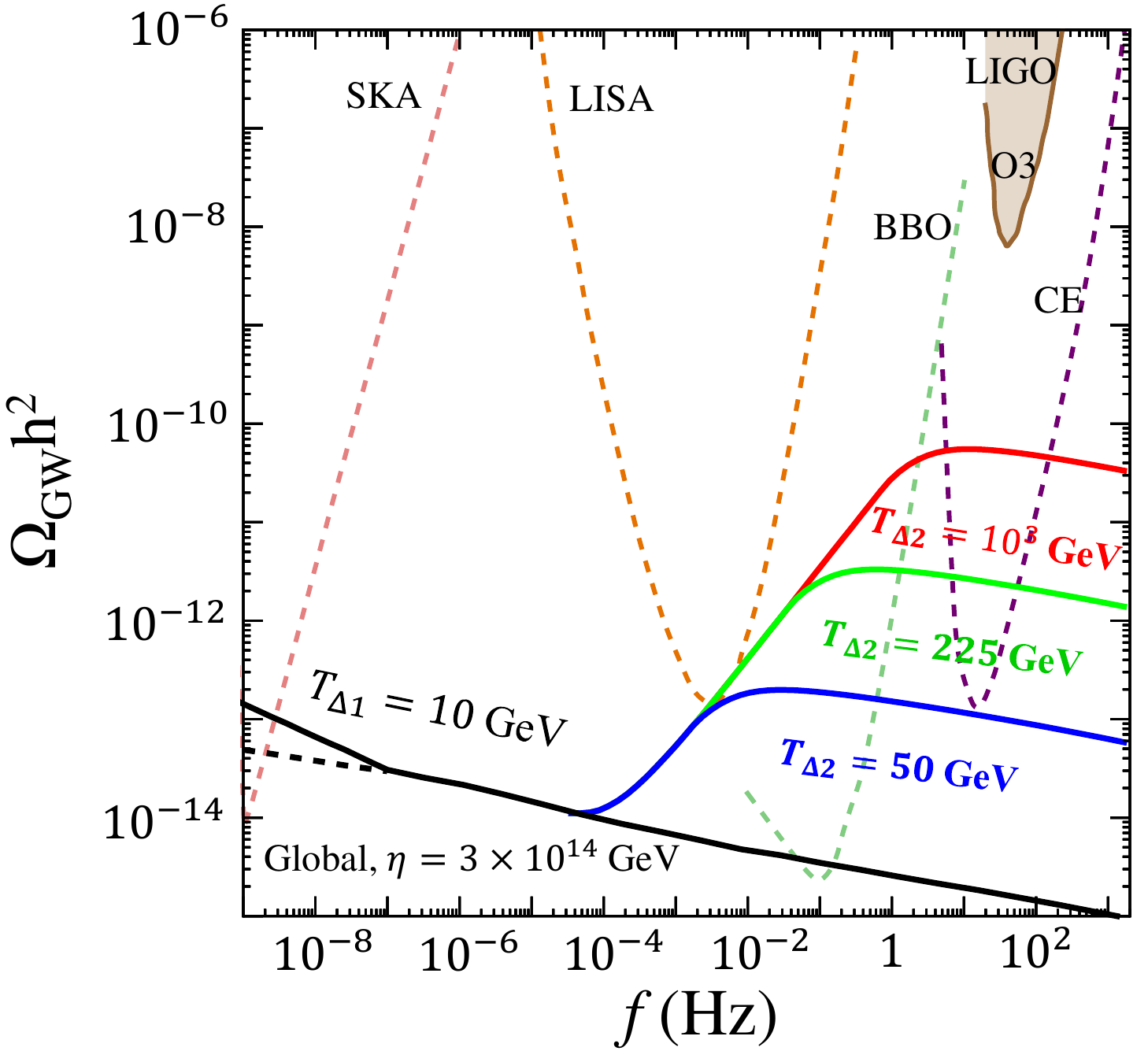} 
\includegraphics[width=0.48\textwidth]{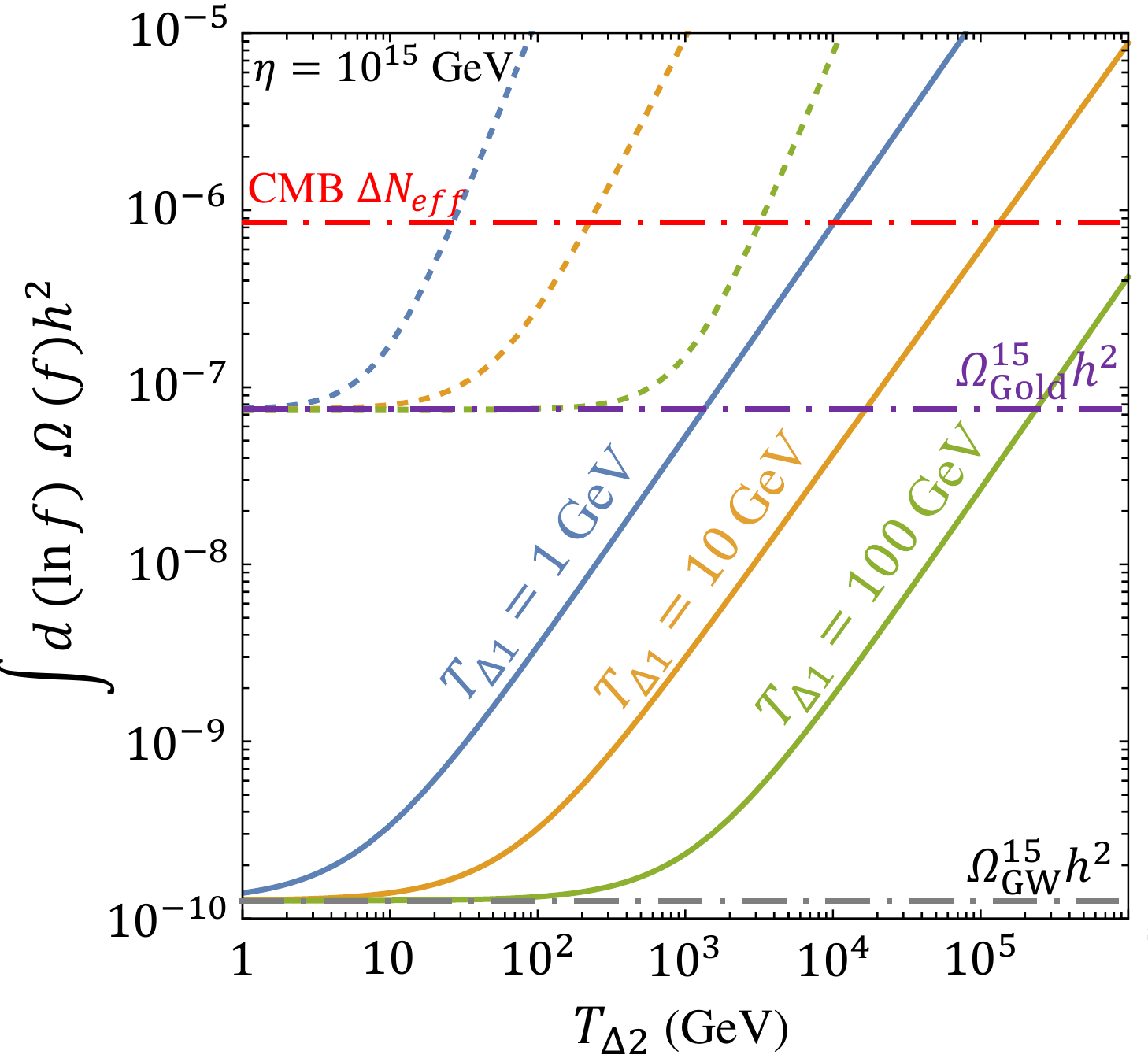} 
\caption{\label{F9-2} Left panel: examples of GW spectra from global strings with two-stage phase transitions including kination: from an early RD era to kination at $T_{\Delta 2}$, and from kination to standard cosmology at $T_{\Delta1}$. Right panel: the relic energy densities of GWs (solid) and Goldstones (dashed) from global strings with varying phase transition temperatures ($T_{\Delta1}$, $T_{\Delta2}$). The red dotted-dashed line shows the CMB bound on extra radiation energy density \cite{Henrot-Versille:2014jua,Aghanim:2018eyx}.} 
\end{figure}

Examples of GW spectrum of such a two-stage transition scenario are shown in the left-panel of Fig.~\ref{F9-2}: $\Omega_{\hbox{\st{GW}}}$ linearly rises with $f$ in a finite frequency range of $f_{\Delta2} > f > f_{\Delta1}$ due to kination, then restores the logarithmically  decreasing behavior in the range of $f > f_{\Delta2}$ (Eq.(\ref{Eq4-2})) during the early RD. The three benchmark cases shown satisfy both LIGO O3 bound and the CMB $\Delta N_{\rm eff}$ bound that we will discuss next. The characteristic frequency $f_{\Delta1}$ corresponding to the later stage of transition at $T_{\Delta 1}$ can be estimated by Eq.(\ref{Eq4-1}). Similarly, based on Eq.(\ref{Eq3-5-2}), the frequency corresponding to the earlier transition at $T_{\Delta 2}$ can be estimated as (applying $\rho\propto a^{-6}$ for kination)
\begin{align}
f_{\Delta 2} = \left( \frac{T_{\Delta 2}}{T_{\Delta 1}} \right)^2 f_{\Delta 1}.
\end{align}

Now we consider the implication of the CMB $\Delta N_{\rm eff}$ bound on additional relic radiation for this kination example. We assume the equation of state of the Goldstones emitted from global strings is radiation-like. As suggested by the sharp rising of GW spectrum shown in Fig.~\ref{F9-2} in the presence of a kination phase, the relic radiation energy densities of Goldstone and GW from the string network are dominated by the emission during the kination epoch, $T_{\Delta 1}<T<T_{\Delta 2}$, which can be roughly estimated as
\begin{align}
\label{Eq: Goldstone Short Kination}
\Omega_{\hbox{\st{Gold}}} h^2 & \,  \sim  \Bigg\{8.0\times 10^{-9} \left( \frac{T_{\Delta1}}{\hbox{GeV}} \right)^{-1.6} \left( \frac{T_{\Delta2}}{\hbox{GeV}} \right)^{1.5}+ \Omega_{\hbox{\st{Gold}}}^{15}h^2 \Bigg\} \left( \frac{\eta}{ 10^{15}\,\hbox{GeV}} \right)^{2} ,\\
\label{Eq4-42}
\Omega_{\hbox{\st{GW}}} h^2 & \, \sim \Bigg\{  1.3\times 10^{-11} \left( \frac{T_{\Delta1}}{\hbox{GeV}} \right)^{-1.37} \left( \frac{T_{\Delta2}}{\hbox{GeV}} \right)^{1.2} + \Omega_{\hbox{\st{GW}}}^{15}h^2 \Bigg\} \left( \frac{\eta}{ 10^{15}\,\hbox{GeV}} \right)^{4} ,
\end{align}
where $\Omega_{\hbox{\st{\{GW,Gold\}}}}$ are defined in Eq.(\ref{Eq3-11}), and numerically computed/fitted based on Eq.(\ref{Eq3-6}). We also defined the reference values with $\eta = 10^{15}\,$GeV assuming the standard cosmology:
\begin{align}
\Omega_{\hbox{\st{Gold}}}^{15}h^2 \simeq 7.5\times 10^{-8}, \;\;\;\;\;\;\;\;\;\;\;\Omega_{\hbox{\st{GW}}}^{15}h^2 \simeq 1.3 \times 10^{-10}.
\end{align}
As discussed in Sec.~\ref{sec:RDGG}, the emission of radiation-like Goldstone dominates the bound. The right panel of Fig.~\ref{F9-2} illustrates three viable benchmark scenarios assuming $\eta = 10^{15}\,$GeV, parametrized by $T_{\Delta1}$, $T_{\Delta2}$: with $T_{\Delta 2} =25\,$GeV, $180\,$GeV, and $3150\,$GeV, the CMB $\Delta N_{\rm eff}$ bound requires $T_{\Delta 1} \gtrsim 1\,$GeV, $10\,$GeV, and $ 100\,$GeV, respectively. 

\subsection{Probing new degrees of freedom}
Many BSM theories involve new particles that are relativistic and in thermal equilibrium in the early Universe, e.g. in many potential solutions to the electroweak hierarchy problem \cite{Chacko:2005pe,Arkani-Hamed:2016rle,Graham:2009gy,Graham:2015cka}, and theories of dark sectors \cite{Feng:2008mu,Adshead:2016xxj,Strassler:2006im,Hodges:1993yb,Kolb:1985bf,Brust:2017nmv,Baumann:2015rya,Chacko:2015noa,Brust:2013ova,Kaplan:2015fuy,Foot:2014mia}. These particles would contribute to the effective number of relativistic degrees of freedom (DOFs) in energy, $g_*$, and in entropy, $g_{*S}$, in the high $T$ Universe, but can generally be out of reach of available probes such as by the LHC or CMB experiments due to heavy masses or feeble interactions with the SM. The methodology for calculating the effect of new DOFs on the SGWB spectrum of NG strings was introduced in \cite{Cui:2018}. In this work, we briefly review the method and apply it to obtain results in the case of global strings.

\begin{figure}[t]
\centering 
\includegraphics[width=0.78\textwidth]{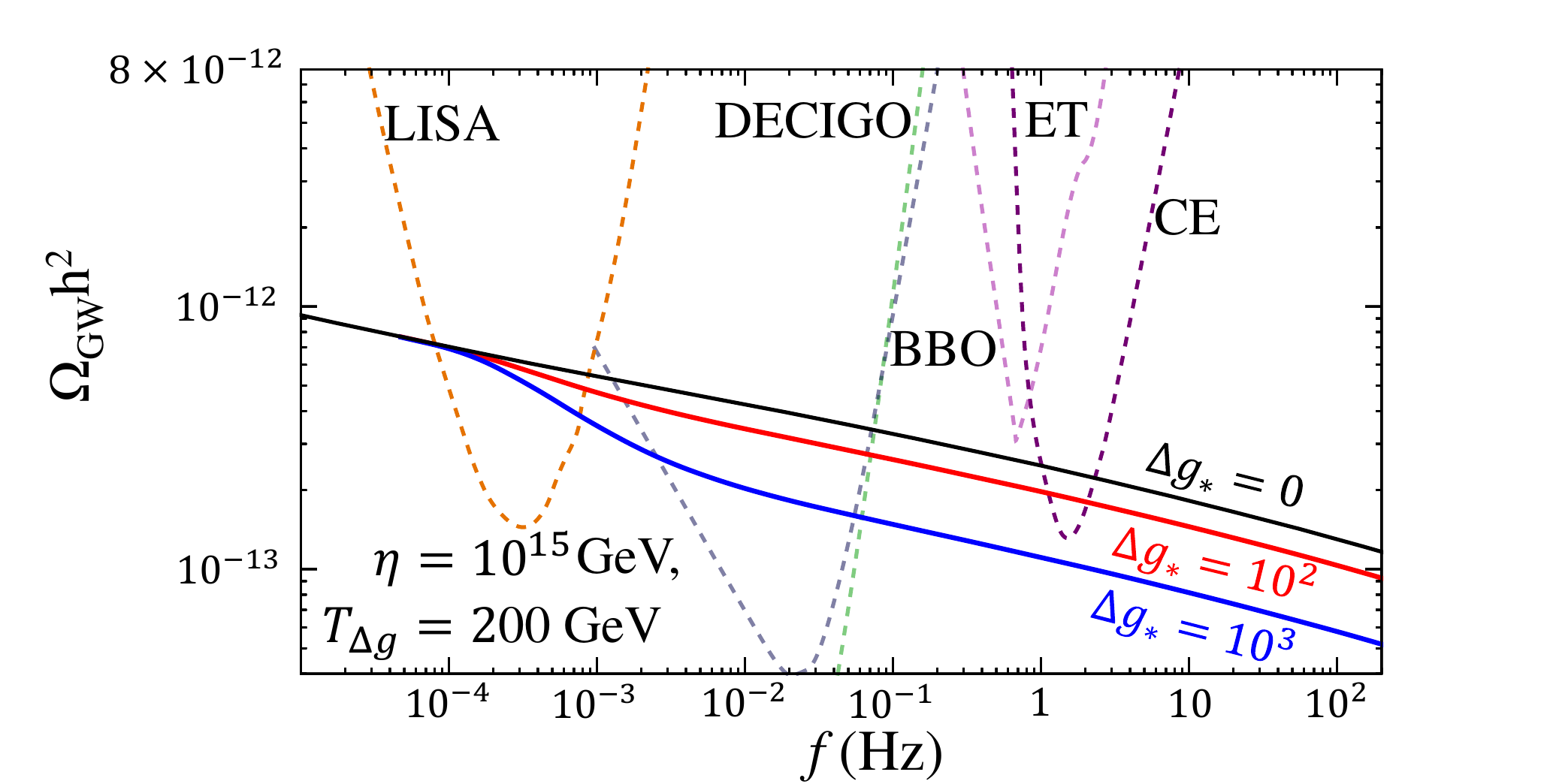} 
\caption{\label{F10} Modification to the GW spectrum from a global string network due to an increase in the number of relativistic degrees of freedom above $T_{\Delta g} = 200\,$GeV. In the example shown, $\eta = 10^{15}\,$GeV, $\alpha = 0.1$, and $\Delta g_* =0, 10^2, 10^3$ (shown in black, red, and blue, respectively). The relevant experimental sensitivities are also shown.}
\end{figure}

We illustrate the effect of new massive DOFs on the string GW spectrum without referring to the details of the underlying theory. We model the change in the number of DOF with the following assumption where $g_*$ rapidly decreases as $T$ drops below a mass threshold $T_{\Delta g}$ \cite{Cui:2018}:  
\begin{align}
\label{Eq4-4}
g_*(T) = g_*^{\hbox{\st{SM}}}(T) + \frac{\Delta g_*}{2} \left[ 1 + \tanh\left( 10 \frac{T-T_{\Delta g}}{T_{\Delta g}}  \right) \right] \simeq \left\{       
\begin{aligned}
&g_*^{\hbox{\st{SM}}}(T) \;\;\;\;\;\;\;\; \;\;\;\;\; ; T< T_{\Delta g} \\
&g_*^{\hbox{\st{SM}}}(T) + \Delta g_*\;\; ; T> T_{\Delta g} 
\end{aligned}
\right.
\end{align}
To numerically demonstrate the effect, we choose the well-motivated scenario, where $T_{\Delta g}$ is of weak scale, which may be motivated from solutions to the Hierarchy Problem. In particular, in Fig.~\ref{F10} we choose the benchmark values of $T_{\Delta g} = 200\,$GeV, $\Delta g_* = 0, 10^2, 10^3$, and assume $g_* \simeq g_{*S}$. As can be seen, relative to the prediction with SM DOFs only, the spectrum falls towards higher $f$ starting from a frequency $f_{\Delta g}$ that agrees with the prediction by the $f$-$T$ relation in the RD era (see Eq.(\ref{Eq4-1})). 

Such an effect can be understood by analytical estimates following \cite{Cui:2018}. Deep in the RD regime, the Hubble rate and the corresponding time depend on $g_*$ in the following way:
\begin{align}
\label{Eq4-5}
H \simeq \sqrt{\Delta _R \Omega_R} H_0 a^{-2}, \;\;\;\;\;\;\; t\simeq \frac{a^2}{2\sqrt{\Delta_R \Omega_R}},
\end{align}
with
\begin{align}
\label{Eq: Delta_R_a}
\Delta_R(a) = \frac{g_*(a)}{g_*^0}\left( \frac{g_{*S}^0}{g_{*S}(a)} \right)^{4/3},
\end{align}
where $H_0$ is the current Hubble constant, and $\Omega_R$ is the radiation energy relic density observed today. Note that $\Delta_R(a)$ is simply a variational form of $\Delta_R(f)$ as defined in Eq.~(\ref{Eq: Delta_R}). Applying this simplification in Eq.(\ref{Eq3-6}), we have
\begin{align}
\label{Eq4-7}
\Omega_{\hbox{\st{GW}}} \left( f \gg f_{\Delta g} \right) \simeq \Omega_{\hbox{\st{GW}}}^{\hbox{\st{SM}}} (f) \left( \frac{g_*^{\hbox{\st{SM}}}}{g_*^{\hbox{\st{SM}}} + \Delta g_*} \right)^{1/3},
\end{align}
where $\Omega_{\hbox{\st{GW}}}^{\hbox{\st{SM}}} (f)$ indicates the amplitude with SM DOFs only. Eq.(\ref{Eq4-7}) clearly shows that the overall amplitude of the high $f$ tail ($f>f_{\Delta g}$) of $\Omega_{\hbox{\st{GW}}}$ decreases with the presence of additional DOFs, agreeing with numerical findings.

\section{Discussion}\label{sec:Diss}

\subsection{Sensitivity to the loop size parameter $\alpha$ and its distribution}\label{Sec:4-5}
As introduced in Sec.~\ref{sec:DoGSL}, throughout our work we have used $\alpha \simeq 0.1$ as the peak value of loop sizes at their formation time, which is inspired by results from NG string simulations \cite{Blanco-Pillado:2017oxo,Blanco-Pillado:2013qja}. However, there are still uncertainties about loop distribution  for global strings. To investigate how such uncertainties may impact the predicted GW spectrum, in this subsection we consider two alternative scenarios of loop distribution: 1.~varying $\alpha$ for the peak value, and 2.~a log uniform distribution of loops as suggested in \cite{Gorghetto:2018myk}. 

\noindent\textit{Alternative-1: varying $\alpha$ for the peak value.}
\begin{figure}[t]
\centering 
\includegraphics[width=0.47\textwidth]{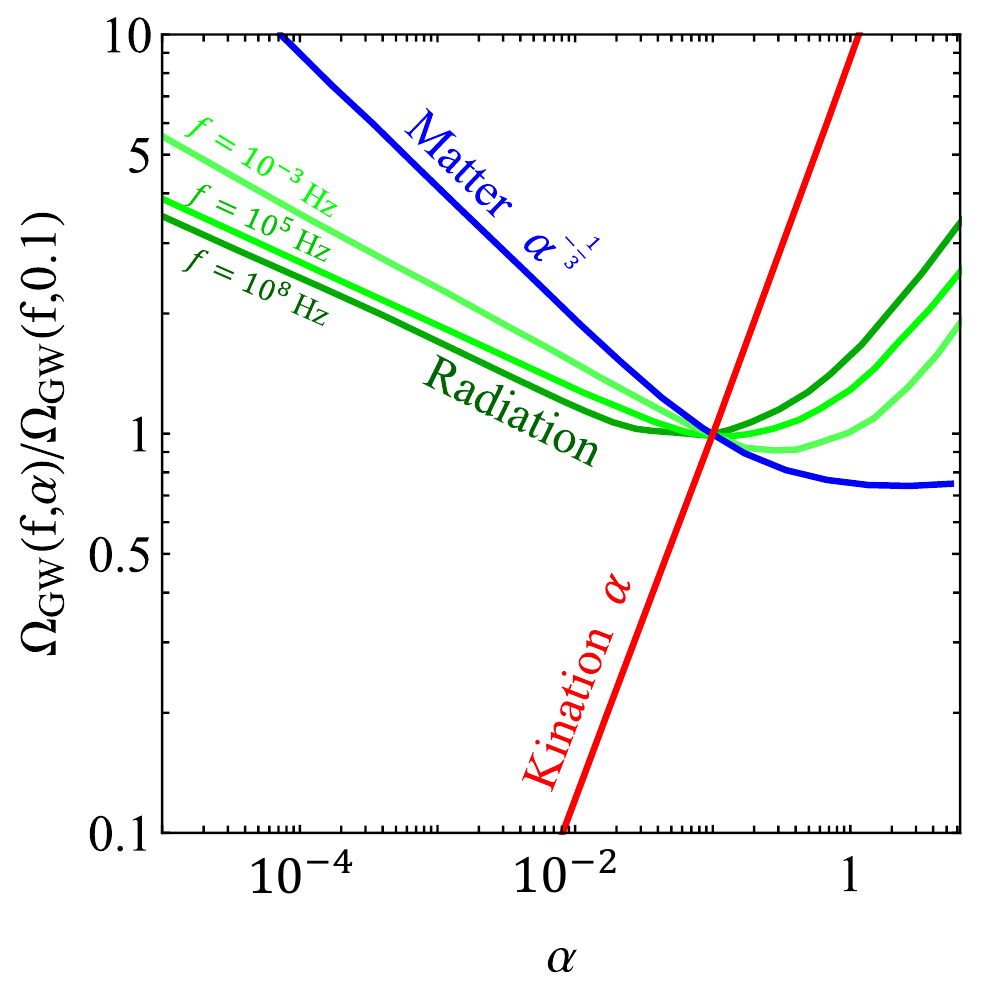} 
\includegraphics[width=0.52\textwidth]{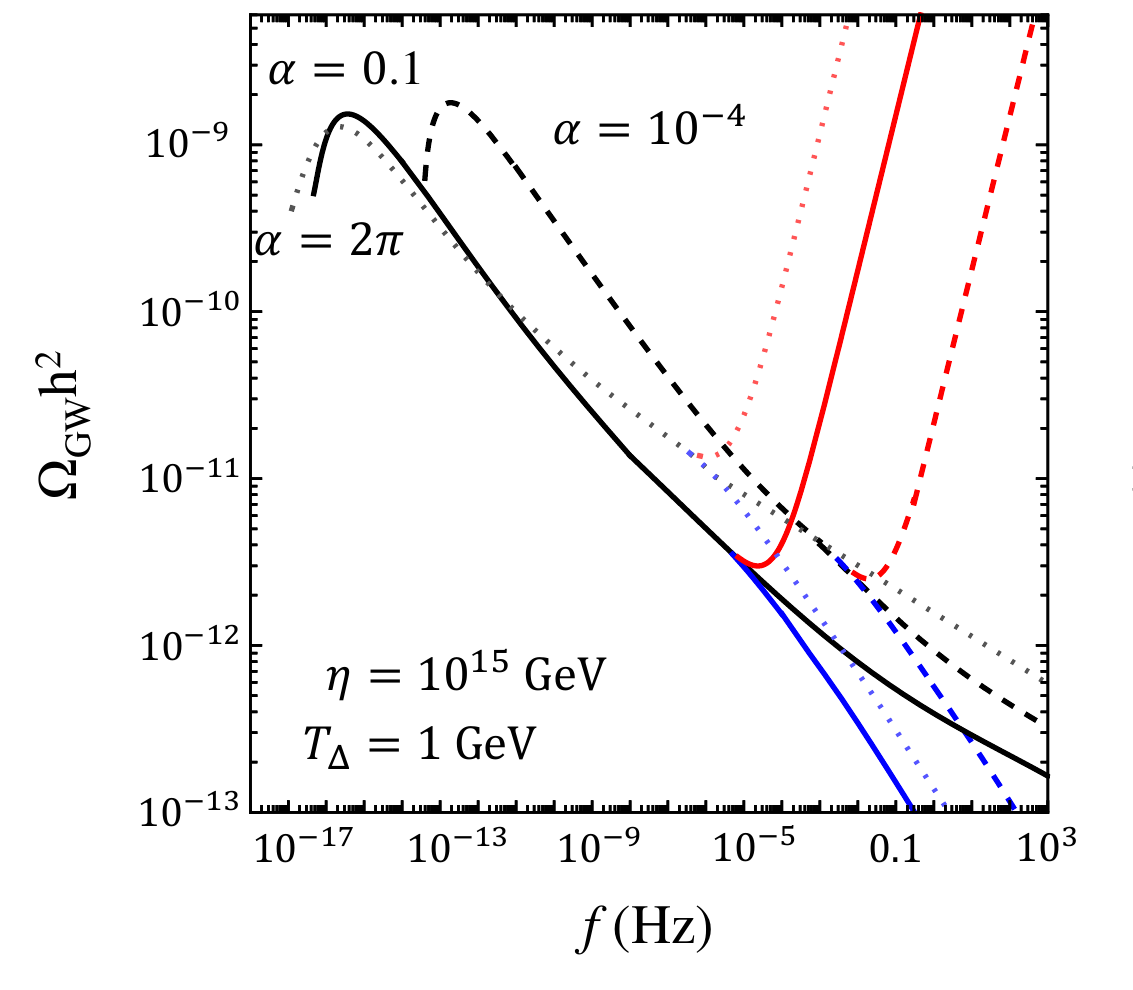} 
\caption{\label{F11} Left panel: $\Omega_{\rm GW}(f, \alpha)$ normalized by the prediction with $\alpha = 0.1$, varying $\alpha$ in the range of $10^{-5}\leq \alpha \leq 10$, $\eta = 10^{15}\,$GeV for different background cosmologies (for MD and kination, the departure from standard cosmology is assumed to occur at $T_\Delta = 1\,$GeV). The green lines show the results with radiation dominated epoch with varying $f=10^{-2}, \; 10^{-5}, \; 10^{-8}\,$Hz, and the red (blue) line shows the results for kination (matter) domination which are insensitive to $f$. Right panel: GW frequency spectra with varying loop size $\alpha$ (dotted: $\alpha=2\pi$, solid: $\alpha =0.1$, dashed: $\alpha=10^{-4}$) with various background cosmologies: standard cosmology (black), kination (red) and EMD (blue)--another way of illustration with the same choices of $\eta$, $T_\Delta$ as in the left panel. }
\end{figure}

The analysis with different $\alpha$ values is straightforward with our formulations in Sec.~\ref{sec:GWSGCS}. In the left panel of Fig.~\ref{F11} we show the $\alpha$ dependence of $\Omega_{\rm GW}(f)$ for specific $f$'s normalized by the prediction with $\alpha = 0.1$ (the benchmark choice used in earlier sections) with different background cosmologies, assuming $\eta=10^{15}\,$GeV. As shown, RD, MD and kination dominated eras have different dependencies on $\alpha$, which are insensitive to $f$ for the cases of MD and kination. The $\alpha$ dependence can be discussed in two distinct regions. Firstly, in the range of $\alpha < \kappa \sim 0.11$, the loop lifetime is shorter than a Hubble time, and thus the analysis and discussion in the previous sections can apply: the 2nd line in Eq.(\ref{Eq5-1}) explains the result for RD. As the loops decay quickly after formation, for a given GW frequency observed today, on average scenarios with smaller $\alpha$ would be associated with larger $N$ at the emission time in order to allow for a longer period of redshifting. Larger $N$ corresponds to higher string tension and thus  larger string energy density available for GW production. In addition, with smaller $\alpha$ the loops emit GW with higher frequency and thus the spectrum blue shifts as shown in the right panel of Fig.~\ref{F11}. Given these two effects, the GW spectrum appears to increase in amplitude as $\alpha$ decreases in this regime. We find that in a kination epoch the spectrum linearly increases with $\alpha$, while in matter domination $\Omega_{\rm GW}(f) \propto \alpha^{-1/3}$. In the other region of $\alpha > \kappa$, the loops are long-lived, and thus the spectrum in RD agrees with the NG string case, which gives $\Omega_{\rm GW}(f) \propto \alpha^{1/2}$ \cite{Cui:2018}. In this large $\alpha$ region, $\Omega_{\rm GW}(f)$ still linearly increases with $\alpha$ in kination, while becomes approximately $\alpha$ independent in MD.
\begin{figure}[t]
\centering 
\includegraphics[width=0.78\textwidth]{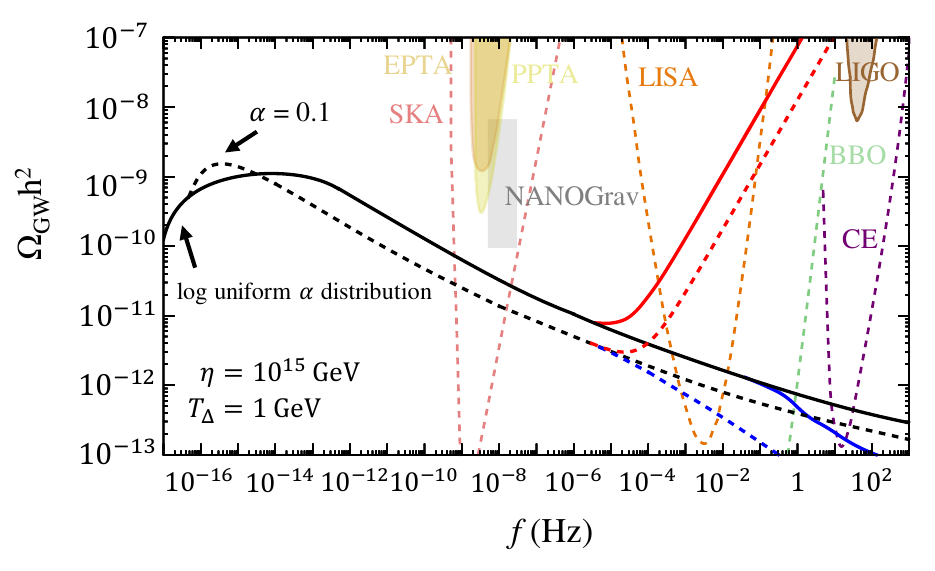} 
\caption{\label{F12} Solid lines: GW spectrum with a logarithmic uniform loop size distribution (Eq.(\ref{Eq4-4})) for different cosmology backgrounds; dashed lines: results with a monotonous $\alpha = 0.1$ as applied in previous sections (for comparison). For the cases with kination or EMD, the departure from standard cosmology is assumed to occur at $T_\Delta =1\,$GeV. }
\end{figure}

\noindent\textit{Alternative-2: A log uniform distribution.}\\
Fig.~5 of the recent global string simulation \cite{Gorghetto:2018myk} suggests a logarithmic uniform distribution of the size of string loops at formation time, which is very different from the nearly monotonous $\alpha$ that we have assumed inspired by NG string simulation. While this hint of log uniform distribution is yet to be further tested, we consider how this variation can impact the prediction for GW signals. A log uniform distribution indicates that at formation time the loop number density $dn(\ell)/d\ell$ at size $\ell$ follows $dn(\ell)/d(\log\ell)\sim\rm const$, or $dn(\ell)/d\ell\propto1/\ell$. Our method of calculating SGWB signal with a monotonous loop formation size $\alpha$ as shown in Eq.(\ref{Eq3-6}) can be adapted to this alternative distribution by replacing $\mathcal{F}_\alpha/\alpha...$ in Eq.(\ref{Eq3-6}) (the ``..." part represent other parts in the formula for computing GWs) with a sum over thinly sliced loop sizes in the range of $\pi/\eta<\ell<\pi/H$:
\begin{align}
\label{Eq: fa summation}
\sum_\alpha \frac{\mathcal{F}_\alpha}{\alpha}... = \lim_{n\to \infty} \sum_{x=0}^n \frac{1}{n} \left( \frac{1}{e^{-\frac{x}{n} \delta + y} } \right)... = \frac{1}{\delta} \int_{y}^{\delta-y} e^x dx... = \frac{1}{\delta} \int_{\alpha_1}^{\alpha_0} \frac{1}{\alpha^2} d \alpha...,
\end{align}
where we have applied $\mathcal{F}_\alpha = \frac{1}{n}$ and $\alpha = e^{-x}$ to implement the log uniform distribution, and taken the continuous limit to get the second equality. The parameters $y,~\delta,~\alpha_0,~\alpha_1$ are introduced to rewrite the integration limits in more convenient forms: $\ell_{\rm max} \sim \pi/H \equiv \alpha_0 t \equiv e^{y} t$, and $\ell_{\rm min} \sim \pi/\eta  \equiv \alpha_1 t \equiv e^{-\delta+y} t$. However, in our numerical calculation we found that including small loops down to the scale of $\pi/\eta$ leads to very large $\Omega_{\hbox{\st{GW}}}(f)\gg 1$ in certain $f$ range as a consequence of energy conservation. Therefore, we assume a lower cutoff of $\alpha$ at $\alpha \sim10^{-4}$.
The exact value of small scale cutoff on $\alpha$ is not essential for our study here, as our purpose is to simply show an example of how a log uniform distribution can alter the GW spectrum. 

In Fig.~\ref{F12} we show the GW spectrum predicted with the assumed log uniform distribution for different cosmology scenarios. We find that by summing over the loop sizes in the range of $10^{-4}\leq \alpha \leq 2 \pi$, the GW amplitude is generally increased over many decades in the frequency range except around the cutoff around $f_0\sim 10^{-16}\,$Hz. Due to the inclusion of larger loops up to $\alpha = 2\pi$ in the distribution, the low frequency cutoff extends to $\sim 2/(2\pi t_0)$. 

\subsection{Sensitivity to the loop radiation parameter $\Gamma$ and $\Gamma_a$}\label{sec:Gamma}
While we chose motivated benchmark values of loop radiation parameters $\Gamma$ and $\Gamma_a$ in our main studies, we acknowledge that there are still uncertainties around these values. Here we investigate how the GW signal would change by varying $\Gamma$ and $\Gamma_a$. Considering energy conservation law and energy loss rates in Eq.(\ref{Eq3-1}), naively, we expect the GW density to depend on $\Gamma, \Gamma_a$ simply as $\Omega_{\hbox{\st{GW}}} \propto \frac{\Gamma G\mu^2}{\Gamma_a \eta^2}$. However, such dependencies can be more complex as the redshift-related  factors $a(\tilde{t})$ and $t_i^{(k)}$ in Eq.(\ref{Eq3-6}) also depend on $\Gamma_a$, $\Gamma$. 
 \begin{figure}[t]
\centering
\includegraphics[width=0.8\textwidth]{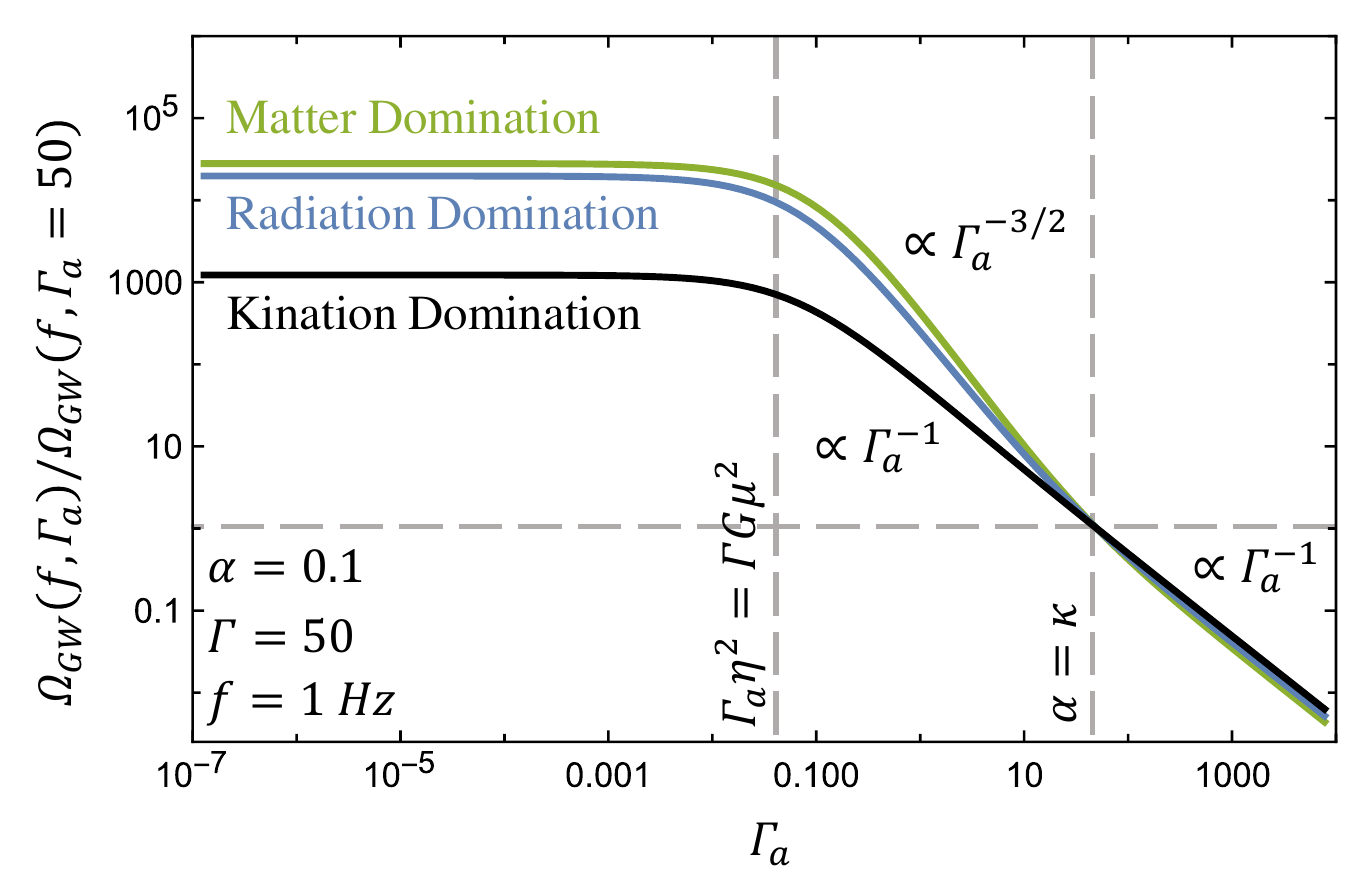} 
\caption{\label{Fig:Gamma-Dependence} An illustration of $\Omega_{\rm GW}(f)$ for varying $\Gamma_a$ (the Goldstone radiation parameter), normalized to the results with $\Gamma_a=50$ (the benchmark value used in earlier sections). We fix other parameters as: $\Gamma = 50$, $\eta = 10^{15}\,$GeV, $f=1\,$Hz. The results for different background cosmologies are shown in different colors. The three regions as discussed in the text are divided by the vertical dashed lines.}

\end{figure}
In Fig.~\ref{Fig:Gamma-Dependence} we illustrate the possibilities for the $\Gamma_a$ dependence of $\Omega_{\hbox{\st{GW}}}(f)$ based on numerical results (fixing $f = 1\,$Hz and $\eta = 10^{15}\,$GeV and $\Gamma=50$ for example). $\Gamma$ dependence is simpler, linear as naively expected, unless GW becomes the dominant radiation mode ($\Gamma_a\ll\Gamma$). We will show the $\Gamma$ dependence explicitly in the following formulae/discussion. As can be seen in Fig.~\ref{Fig:Gamma-Dependence} there are three distinct regions in the $\Omega_{\hbox{\st{GW}}}(f)-\Gamma_a$ relation, which we can understand analytically as follows: \\
\noindent$\bullet$ Large $\Gamma_a$, such that loops decay within a Hubble time after formation, driven by strong Goldstone emission. In this region $\alpha < \kappa$, where $\kappa \equiv \Gamma_a/(2\pi N)$ (Eq.(\ref{Eq:ti k})), and we can estimate with $N \sim 70$ for relevant observations. 
The $\Gamma_a$ term thus dominates both numerator and denominator of Eq.(\ref{Eq:ti k}), which implies that both $a(\tilde{t})$ and $t_i^{(k)}$ in Eq.(\ref{Eq3-6}) are insensitive to $\Gamma_a$. Therefore, $\Omega_{\hbox{\st{GW}}}(f)$ in Eq.(\ref{Eq3-6}) depends on $\Gamma, \Gamma_a$ as 
\begin{align}
\Omega_{\hbox{\st{GW}}} (f) \propto \frac{\Gamma}{\Gamma_a}. 
\end{align}

\noindent$\bullet$ Medium size $\Gamma_a$, such that loops survive beyond a Hubble time after formation, while Goldstone radiation still dominates over GWs. In this region, $\alpha > \kappa> \Gamma G \mu$, and thus redshift factors $a(\tilde{t})$ and $t_i^{(k)}$ in Eq.(\ref{Eq3-6}) depend on $\Gamma_a$. 
By fitting numerical results we find the following relations which depend on background cosmologies:
\begin{align}
\Omega_{\hbox{\st{GW}}} (f)\propto\left\{           
\begin{aligned}
&\; \frac{\Gamma}{\Gamma_a^{3/2}}, \;\;\;\;\;\;\hbox{for RD and EMD}, \;\;\;\;\;\\
&\; \frac{\Gamma}{\Gamma_a}, \;\;\;\;\;\;\;\;\;\hbox{for Kination}. \;\;\;\;\; 
\end{aligned}
\right.
\end{align}
As discussed in Sec.~\ref{sec:PNPOCE}, the GW frequency spectrum with an EMD is dominated by loop radiation during the later radiation domination era. Consequently, the $\Omega_{\rm GW}(f)$-$\Gamma_a$ relation is approximately the same as RD for the benchmark frequency $f=1\,$Hz. 

\noindent$\bullet$ Small $\Gamma_a$, such that $\Gamma_a \lesssim \Gamma G \mu^2 /\eta^2$ (i.e. $\kappa < \Gamma G \mu$, and the Goldstone emission term in Eq.(\ref{Eq3-1}) becomes negligible relative to the GW radiation). Given the hierarchy between the Planck mass and the viable $\eta$ value considering the relevant constraints, this scenario is only possible for very small $\Gamma_a\ll\Gamma$. In this case, GW radiation would become the dominant energy loss mechanism and $\Omega_{\hbox{\st{GW}}} (f)$ would increase as, $\Gamma^{-1/2}$ which agrees with the related result for NG strings \cite{Cui:2018}. 

\subsection{Non-scaling solution}\label{Sec:Non-scaling solution} 
In this subsection, we consider the impact of possible non-scaling solutions on the GW signals. The violation of the scaling properties in the case of global strings were found in some recent simulation studies \cite{Vaquero:2018tib,Klaer:2019fxc,Klaer:2017qhr,Kawasaki:2018bzv,Buschmann:2019icd,Fleury:2015aca,Gorghetto:2018myk,Gorghetto:2020qws}. This suggests that the attractor solution of the average number of strings per Hubble patch, $\xi$, logarithmic growing with $N$. Note that in most of these studies the non-scaling behavior is found in the low $N$ regime which is within direct reach of current simulations, and whether such a behavior can apply to large $N$ still needs to be investigated. For example, Ref.~\cite{Hindmarsh:2021zkt} pointed out that the discrepancy in literature could be due to the different interpretations of the initial growth of $\xi$ by different groups: the data set in the earlier study Ref.~\cite{Gorghetto:2018myk} may be too close to the string formation, and thus is sensitive to initial condition. They further showed that the solution should converge to the constant estimation of $\xi \simeq 1.19$ as given in \cite{Hindmarsh:2021vih,Hindmarsh:2019csc}. We found that by assuming the same VOS parameters, such a constant $\xi$ would decrease GW amplitude to about $15\%$ due to $C_{\rm eff} \propto \xi^{3/2}$ dependence (see Eq.(\ref{Eq: Ceff})).

As earlier shown in Fig.~\ref{F1}, the VOS model can be consistent with the non-scaling solution Eq.(\ref{Eq2-6-2}) (or Eq.(\ref{Eq: Xi}) below) within the range of low $N$, $3 \lesssim N \lesssim 7$, then predicts $\xi \sim$ const.~for larger $N$. Nevertheless, it is intriguing to see how the GW spectrum would change if such a behavior does sustain throughout the evolution history of the string network, and whether/how a variation to the original analytical VOS model may match this behavior. We will focus on the following two examples and then comment on other possibilities, and in both cases we adopt the non-scaling solution as suggested in simulations \cite{Klaer:2019fxc,Gorghetto:2021fsn,Gorghetto:2020qws}
\begin{align}
\label{Eq: Xi}
\xi = 0.24(2) N + 0.2,
\end{align}
where $N \equiv \hbox{ln}(\eta/H(t))$. We consider $\xi$ taking the above non-scaling form in both examples that we will discuss next, and adopt the relevant parameters from the VOS model for the GW calculations (Eqs.(\ref{Eq2-7},\ref{Eq2-8},\ref{Eq: Ceff})). 

\begin{figure}[t]
\centering
\includegraphics[width=0.8\textwidth]{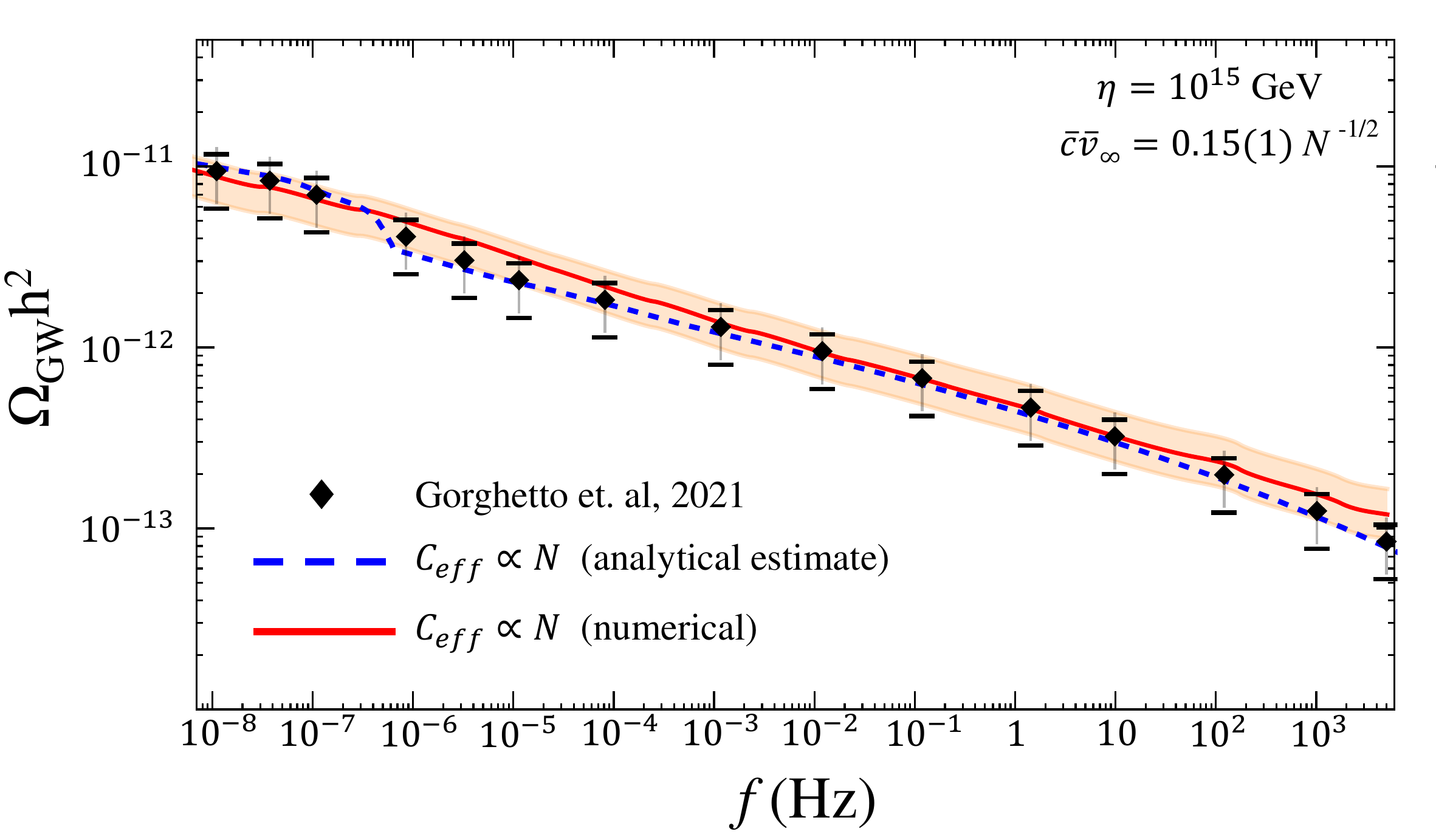} 
\caption{\label{F14} GW spectrum in the radiation dominated epoch assuming a non-scaling solution ($\xi = 0.24(2)N + 0.2$): a comparison between the result with our assumption/method and that obtained in the recent simulation work \cite{Gorghetto:2021fsn}. The data points with error bars are taken from \cite{Gorghetto:2021fsn}. The blue dashed curve is based on our analytical estimate Eq.(\ref{Eq: Comparison Marco}). The red curves with shadowed uncertainty band is based on our numerical calculation of Eq.(\ref{Eq3-6}) with linear growth of $C_{\hbox{\st{eff}}} \propto N$. Further details are given in the main text. }
\end{figure}

In the first possibility we consider, in addition to Eq.~\ref{Eq: Xi}, we apply the following benchmark parameters: a constant average velocity of long strings $\bar{v}_\infty \simeq 0.50 \pm 0.04$, and a loop chopping parameter $\bar{c}=0.497$, which we obtained in Sec.~\ref{sec:VOS} based on fitting simulation results (Table.~\ref{Table_Data}). With Eq.(\ref{Eq2-8}), primarily derived based on energy conservation, we find the prediction for effective loop formation parameter $C_{\hbox{\st{eff}}} \propto N^{3/2}$. With this $C_{\hbox{\st{eff}}}$ as an input for Eq.(\ref{Eq3-6}) we computed the GW spectrum, and found that the amplitude is larger than the prediction in \cite{Gorghetto:2021fsn} by a factor of $\mathcal{O}(10-100)$, depending on frequencies.  This discrepancy motivated us to introduce the second scenario which is found to lead to a good agreement with \cite{Gorghetto:2020qws}: while still assuming Eq.~(\ref{Eq: Xi}), this example involves a time-dependent $\bar{c}\bar{v}_\infty$, and consequently a different form of $C_{\hbox{\st{eff}}}$: 
\begin{align}
\label{Eq: C V nonscaling}
\bar{c}\bar{v}_\infty = 0.15(1) N^{-1/2} \;\;\;\;\; \to \;\;\;\;\; C_{\hbox{\st{eff}}} \simeq 0.018(3) N.
\end{align}
Based on the analysis method given in Sec.~\ref{sec:RDGG}, the GW spectrum with the non-scaling solution Eq.(\ref{Eq: C V nonscaling}) in a RD background (the spectrum would be cut off at lower frequencies by a QCD-like phase transition as shown in \cite{Gorghetto:2021fsn}) can be estimated as
\begin{align}
\label{Eq: Comparison Marco}
\Omega_{\hbox{\st{GW}}}h^2 \simeq 2.6\times 10^{-17} \left(\frac{\eta}{10^{15}\,\hbox{GeV}}\right)^{4} \log^4 \left[ \left( \frac{2}{\alpha f}\right)^{2} \frac{\eta}{t_{eq}} \frac{1}{2z_{eq}^2 } \Delta_R^{1/2}(f) \right] \Delta_R(f).
\end{align}
The notable difference between this result and that based on the scaling VOS model solution (Eq.(\ref{Eq5-1})) is the power law index of the $\log$ term (i.e. $\log^4$ vs. $\log^3$), which enhances the GW amplitude by $\mathcal{O}(10)$ for this non-scaling example. The enhancement is due to the increase in loop number density (Eq.~\ref{Eq2-8}). Fig.~\ref{F14} illustrates the GW spectrum predicted with a non-scaling solution where $\xi\propto N$, including a comparison between our results based on a variation to the VOS model (Eq.~(\ref{Eq: C V nonscaling})) and the result in \cite{Gorghetto:2021fsn} based on extrapolating simulation results to large $N$. A good agreement between our second scenario (Eq.(\ref{Eq: C V nonscaling})) and that in \cite{Gorghetto:2021fsn} can be seen in Fig.~\ref{F14}. In particular ,our analytic fit for the GW spectrum (Eq.(\ref{Eq: Comparison Marco})) captures the key $\log^4$ dependence that agrees with \cite{Gorghetto:2021fsn}. This agreement suggests that the extrapolation of the non-scaling solution to large $N$ may be reproduced in a variation to the original VOS model, where the relations $\bar{c}\bar{v}_\infty \propto N^{-1/2}$ and $\xi\propto N$ are realized. This hint may be helpful for future investigations. 

\begin{figure}[t]
\centering
\includegraphics[width=1\textwidth]{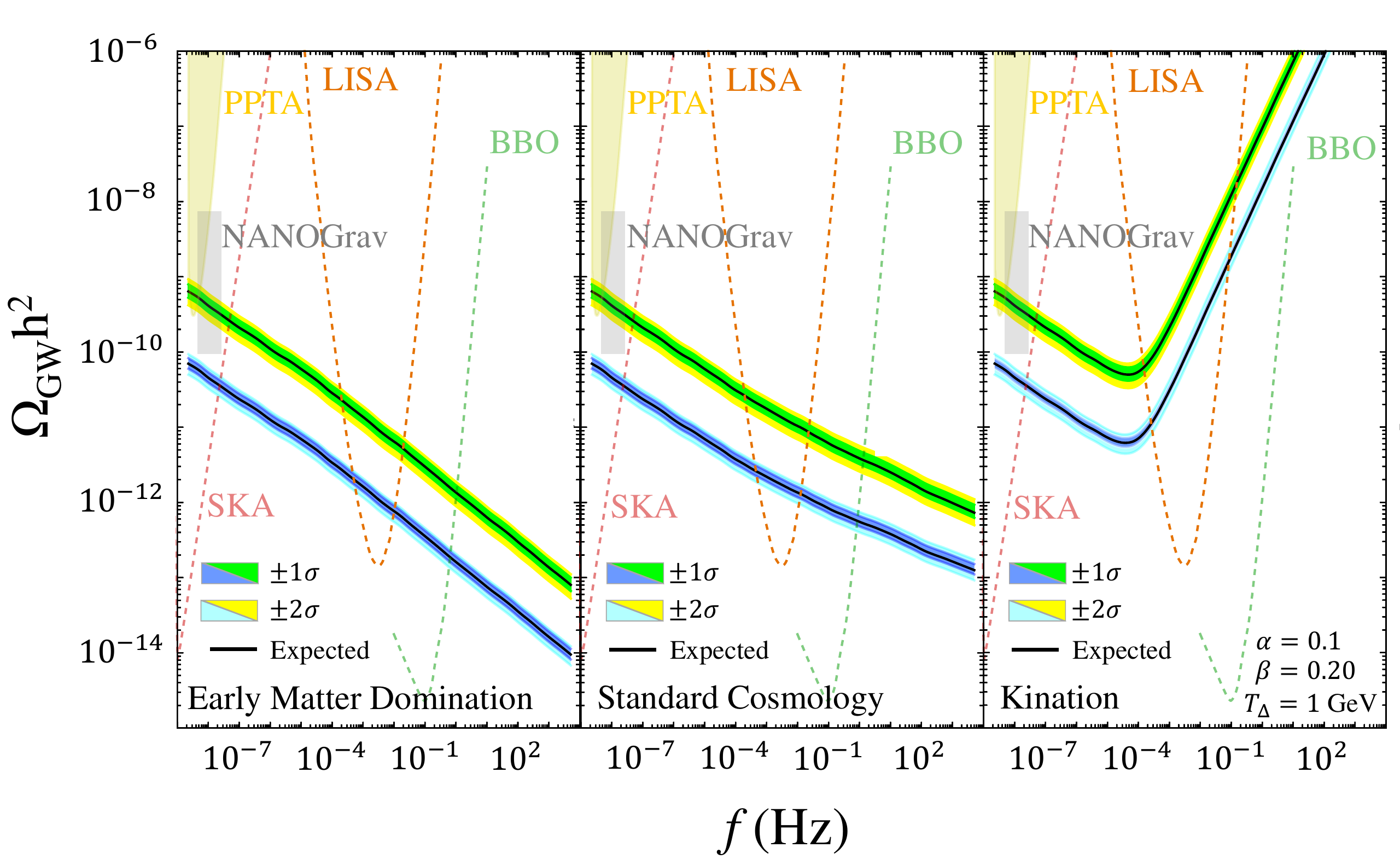} 
\caption{\label{F13} An example of GW spectra from a global string network with a non-scaling solution Eq.(\ref{Eq2-6-2}) in various cosmological backgrounds. Upper lines show the results with $C_{\hbox{\st{eff}}} \propto N$, lower lines show the results with $C_{\hbox{\st{eff}}} \propto N^{3/2}$. Details about the two scenarios of $C_{\hbox{\st{eff}}}$ can be found in the main text. The black lines show the central values, while the yellow(Green) and cyan(Blue) areas represent the 2(1) sigma uncertainty range for the linear growth $\xi = 0.24(2)N + 0.2$. A set of related experimental sensitivities are also shown.}
\end{figure}

As a supplemental discussion, in Fig.~\ref{F13} we illustrate and compare the different predictions of GW spectrum based on the two aforementioned non-scaling scenarios, with various background cosmologies. As shown, in the second scenario (Eq.(\ref{Eq: C V nonscaling})) the GW spectrum amplitude is lowered by $\mathcal{O}(10)$ relative to the first scenario (Eq.(\ref{Eq: Xi})), which is due to the different predictions for loop number density ($C_{\hbox{\st{eff}}}\propto N$ v.s.~$C_{\hbox{\st{eff}}}\propto N^{3/2}$).
In the frequency range of our interest, the result is insensitive to the initial condition dependent parameter $\beta$: as discussed in \cite{Gorghetto:2018myk,Gorghetto:2020qws}, the linearly growing term in Eq.(\ref{Eq: Xi}) would quickly dominate the string network evolution. 

A very different form of non-scaling solution was suggested in another simulation work \cite{Buschmann:2019icd}:
 \begin{align}
\label{Eq: Sadfi2019}
\xi = 2.60 \times \hbox{log}\left( \frac{T_{\hbox{\st{PQ}}}}{T} \right) + 1.27,
\end{align}
where $T_{\hbox{\st{PQ}}} \sim \eta$ is the temperature when the PQ symmetry breaking occurs. The prediction for $\xi$ as in Eq.(\ref{Eq: Sadfi2019}) is significantly larger than non-scaling results from other groups' simulations \cite{Klaer:2019fxc,Klaer:2017qhr,Kawasaki:2018bzv,Gorghetto:2018myk,Gorghetto:2020qws}. As suggested by the authors of \cite{Buschmann:2019icd}, the prediction of Eq.(\ref{Eq: Sadfi2019}) only provides a rough counting for cosmic strings, which may address the discrepancy, while further investigations are needed. We attempted to fit Eq.(\ref{Eq: Sadfi2019}) with a variation of VOS model, but found a rather poor VOS model fit for the 13 data points provided in \cite{Buschmann:2019icd} due to the large value $\xi$ given in Eq.(\ref{Eq: Sadfi2019}), which is inconsistent with other simulation results. Assuming the non-scaling behavior as in form of Eq.(\ref{Eq: Sadfi2019}) sustains till late times, we expect the GW amplitude to be amplified by a factor of $\mathcal{O}(10-100)$ relative to the scaling case due to the larger loop density implied (similar to the case inspired by \cite{Gorghetto:2018myk}).

\subsection{Distinguish from other SGWB sources}
In this subsection, we discuss potential challenges for detecting a SGWB signal from global strings in practice, including astrophysical background and a comparison with other cosmological sources of SGWB.

SGWB from global strings, like other cosmogenic SGWBs, may be contaminated by astrophysical sources of SGWB, e.g. from unresolved binary black hole mergers \cite{TheLIGOScientific:2016wyq,TheLIGOScientific:2017qsa,Abbott:2017gyy,Abbott:2017vtc,Abbott:2017xzg,TheLIGOScientific:2016dpb,Barish:2020vmy}. Progress has been made in recent years to address this important issue of distinguishing a cosmological SGWB from its astrophysical counterpart. The potential solutions include: identify and subtract astrophysical sources using information from future GW detectors with improved resolutions \cite{Abbott:2017xzu,Regimbau:2016ike,Jenkins:2018nty}; optimized statistical analysis beyond the conventional cross-correlation method \cite{Smith:2017vfk,Bartolo:2018qqn,Ginat:2019aed}; utilize spectral information over a wide frequency band \cite{Thrane:2013oya,Baghi:2019eqo,Caprini:2019pxz,Smith:2019wny,Flauger:2020qyi,Barish:2020vmy,Boileau:2020rpg,Schmitz:2020syl,Romano:2016dpx}. Detailed discussions on this subject can be found in e.g. \cite{Cui:2018, Barish:2020vmy,Smith:2019wny}.

Upon detection of a cosmogenic SGWB signal, it is important to analyze and identify the nature of the underlying physics. Global cosmic string is among many motivated new physics sources that can give rise to a SGWB \cite{Caprini:2018mtu,Binetruy:2012ze,Kuroyanagi:2018csn}, for example, primordial inflation \cite{Starobinsky:1979ty,Allen:1987bk} and black hole \cite{Vaskonen:2020lbd}, preheating \cite{Khlebnikov:1997di,Easther:2006gt,Easther:2006vd,GarciaBellido:2007dg}, first-order phase transitions \cite{Witten:1984rs,hogan1986gravitational,kosowsky1992gravitational,Alanne:2019bsm,Schmitz:2020rag}, and other types of topological defects \cite{Gleiser:1998na,Figueroa:2012kw} including local/NG  strings \cite{Vilenkin:1981bx,Vachaspati:1984dz,vachaspati1985gravitational,caldwell1992cosmological,Damour:2004kw,Bevis:2006mj}. A key to distinguishing the various cosmological sources lies in the GW spectral information. For instance, SGWB from a first-order phase transition features a peaky spectrum in frequency associated with specific split power laws, which results from the fact that the GWs were emitted during a specific epoch in the early Universe. In contrast, SGWBs from cosmic strings (both global and NG) feature a rather long (nearly) flat plateau towards high frequencies, due to the continuous emission throughout the cosmic history. We refer to \cite{Cui:2018} for more detail regarding the general comparison of SGWB originated from cosmic strings with other cosmological sources. Here we highlight the prospect of distinguishing SGWB from global strings vs. that from NG strings. As seen in Sec.~\ref{sec:SGSN} and Fig.~\ref{F3} the GW spectrum from global strings has a long tail which logarithmically declines towards high frequencies, whereas the spectrum from NG strings is very close to simple flatness (except for the mild steps due to the change in $g_*$). A main cause of such a difference is the logarithmic time-dependence of the global string tension, Eq.(\ref{Eq: Xi}). The difference would be further amplified if the non-scaling behavior as discussed in Sec.~\ref{Sec:Non-scaling solution} is confirmed to last till late times. In practice, we therefore expect that for global strings, GW searches at lower frequencies such as SKA in general have a better prospect of detection than those at higher frequencies such as LIGO (the prospect also depends on the experimental sensitivities).

In summary, while challenges for experimentally detecting a global string sourced SGWB are present, potentially promising solutions exist and will be further developed in coming years. Using frequency band information is a common potential solution for disentangling a global string signal from both astrophysical background and other cosmological sources, which will be strengthened with a multi-band GW experimental program \cite{Lasky:2015lej,Smith:2019wny,Schmitz:2020syl,Barish:2020vmy}.

\section{Conclusion}\label{sec:con}
Global or axion cosmic strings are well motivated sources of SGWB, and have attracted growing interest in the past few years. In this work, we applied the analytical VOS model in solving the evolution of a global string network over the course of the cosmic history, and illustrated the procedure of calculating the resultant GW signals with great detail. We demonstrated how our VOS model parameters were calibrated by simulation data which are most reliable for the early time of evolution $N \lesssim 7$, and commented on the compatibility between VOS model prediction and simulation results found by various groups. We found that the deviation from the scaling property as found by some simulation studies can be consistent with conventional VOS model prediction in the early regime of $3 \lesssim N \lesssim 7$, but the simple extrapolation of such a non-scaling behavior to large $N$ or late times contradicts the conventional VOS model. Nevertheless, we also investigated how the SGWB signal would alter if the non-scaling does persist to late times, and suggested a possible revision to the VOS model that addresses this difference which can lead to a GW signal prediction consistent with that given in \cite{Gorghetto:2021fsn} based on simulation. While it will take time to resolve the discrepancies among different simulation data sets as well between VOS model prediction and some simulation results, our methodology of analysis and related discussions are timely complements to the literature, and can be further improved/updated in light of future developments.

Our main results are presented following the standard VOS model and analytical calculation of SGWB by summing over harmonic modes and taking into account the significant effect of Goldstone emissions. In light of the recent findings on the important effect of high $k$ modes for NG strings, we summed over $10^5$ modes in all our analysis, leading to updated spectra relative to our earlier results in \cite{Chang:2019mza}. 
We first demonstrated the results assuming standard cosmology, and then considered the possible presence of a non-standard equation of state before BBN, e.g. an EMD or kination, which would lead to a drastic departure from the standard prediction at high $f$ ranges. Since an indefinitely long kination period is subject to strong constraints on additional relic radiation energy density from CMB/BBN data due to a blue-tilted spectrum, we also considered an example where the kination epoch has a finite window of span and is preceded by an early stage of RD. We further demonstrated how the presence of new relativistic degrees of freedom in the early Universe can alter the GW spectrum. We showed the current and projected future sensitivities of GW detectors in detecting global string signals, and found that a detectable signal requires the corresponding spontaneous symmetry breaking scale $\eta>2 \times 10^{14}\,$GeV. Different from NG strings, GW amplitude from global strings is very sensitive to $\eta$ ($\Omega_{\rm GW}\propto \eta^4$). The frequency-time (temperature) relation, which is the foundation for the method of GW cosmic archaeology, takes a very different form for global strings relative to its NG string counterpart. In particular, there is no $G\mu$ dependence in the $f$-$T$ relation and the same $f$ band corresponds to a much earlier emission time for global strings, which enables us to test the standard radiation era up to $T\sim10^8$ GeV. We explained the physics behind the notable differences between SGWB from global strings and from NG strings, where the strong rate of Goldstone emission and the consequent short lifetime of global string loops play an important role.

We further considered how the GW signal based on our baseline assumptions and model choices could vary with alternative possibilities. We studied the effects of different loop distribution patterns, the uncertainty in the radiation parameters ($\Gamma$, $\Gamma_a$), as well as a persisting non-scaling regime during the string network evolution. For example, by adopting the suggested non-scaling solution with $\xi\propto N$ while assuming $\bar{c}\bar{v}_\infty \propto N^{-1/2}$, we found that the predicted GW frequency spectrum (including the $\log^4$ relation) can be consistent with the simulation-based finding in \cite{Gorghetto:2021fsn}. We also briefly discussed the prospect of distinguishing a SGWB sourced by global strings from other cosmogenic sources or astrophysical background. 

It is worth noting the importance of studying GWs from global/axion strings in light of its connection to axion physics (for QCD axion or general axion-like particles (ALPs)). Axion strings are indispensable companions of axion particles when the $U(1)_{\rm PQ}$ symmetry breaking occurs after inflation. The detection of axion particles is being actively pursued, but the prospect is model-dependent due to the uncertainty of the interaction between the SM and the axions. The prospect would be particularly dim in the case of the hidden axion scenario (e.g. motivated by the string axiverse \cite{Arvanitaki:2009fg}) where non-gravitational coupling is absent. Thus, the universal GW signal from axion strings could be the smoking-gun for the underlying axion physics. While our current work focused on the simpler case of pure global strings with massless Goldstones, the methodology and results are relevant for the massive axion case. The GW spectrum from axion topological defects at high frequencies is expected to be dominated by cosmic strings, while at a low frequency corresponding to a QCD(-like) phase transition when the domain wall forms, the signal is expected to change form and die down. The recent work based on extrapolating simulation results to late times shows such a pattern \cite{Gorghetto:2021fsn}. It is intriguing to apply our analytical approaches to the more complex axion scenario including domain wall contributions, which will be pursued in future work. 

\acknowledgments

CC would like to thank Yin Chin Foundation of U.S.A. for its support. YC thanks the Kavli Institute for Theoretical Physics (supported by the National Science Foundation under Grant No. NSF PHY-1748958) for support and hospitality while the work was being completed. The work is supported in part by the US Department of Energy under award number DE-SC0008541. 



\bibliographystyle{JHEPmod}
\bibliography{globalcsGWlong_JHEP}

\end{document}